\documentclass[a4paper,fleqn,usenatbib]{mnras}
\usepackage[T1]{fontenc}
\usepackage{ae,aecompl}
\usepackage{graphicx}
\usepackage{amsmath}
\usepackage{amssymb}
\usepackage{tabularx}

\graphicspath{{Figures/}}

\title[Massive Cluster Evolution]{The redshift evolution of massive galaxy clusters in the MACSIS simulations}
\author[Barnes et al.]
{David J. Barnes$^1$\thanks{Contact e-mail: \href{mailto:david.barnes@manchester.ac.uk}{david.barnes@manchester.ac.uk}}, Scott T. Kay$^1$, Monique A. Henson$^1$, Ian G. McCarthy$^2$, \newauthor Joop Schaye$^3$ and Adrian Jenkins$^4$
\\
$^{1}$Jodrell Bank Centre for Astrophysics, School of Physics and Astronomy, The University of Manchester, Manchester M13 9PL, UK\\
$^{2}$Astrophysics Research Institute, Liverpool John Moores University, 146 Brownlow Hill, Liverpool L3 5RF, UK\\
$^{3}$Leiden Observatory, Leiden University, P.O. Box 9513, 2300 RA Leiden, the Netherlands\\
$^{4}$Institute for Computational Cosmology, Department of Physics, University of Durham, South Road, Durham DH1 3LE, UK
}

\date{Accepted XXX. Received YYY; in original form ZZZ}
\pubyear{2016}

\begin{document}
%\tracingall

\label{firstpage}
\pagerange{\pageref{firstpage}--\pageref{lastpage}}
\maketitle

\begin{abstract}
We present the MAssive ClusterS and Intercluster Structures (MACSIS) project, a suite of 390 clusters simulated with baryonic physics that yields realistic massive galaxy clusters capable of matching a wide range of observed properties. MACSIS extends the recent BAHAMAS simulation to higher masses, enabling robust predictions for the redshift evolution of cluster properties and an assessment of the effect of selecting only the hottest systems. We study the observable-mass scaling relations and the X-ray luminosity-temperature relation over the complete observed cluster mass range. As expected, we find the slope of these scaling relations and the evolution of their normalization with redshift departs significantly from the self-similar predictions. However, for a sample of hot clusters with core-excised temperatures $k_{\rm{B}}T\geq5\,\rm{keV}$ the normalization and slope of the observable-mass relations and their evolution are significantly closer to self-similar. The exception is the temperature-mass relation, for which the increased importance of non-thermal pressure support and biased X-ray temperatures leads to a greater departure from self-similarity in the hottest systems. As a consequence, these also affect the slope and evolution of the normalization in the luminosity-temperature relation. The median hot gas profiles show good agreement with observational data at $z=0$ and $z=1$, with their evolution again departing significantly from the self-similar prediction. However, selecting a hot sample of clusters yields profiles that evolve significantly closer to the self-similar prediction. In conclusion, our results show that understanding the selection function is vital for robust calibration of cluster properties with mass and redshift.
\end{abstract}


\section{Introduction}
\label{sec:intro}

Galaxy clusters form from large primordial density fluctuations that have collapsed and virialised by the present epoch, with more massive clusters forming from larger and rarer fluctuations. This makes them especially sensitive to fundamental cosmological parameters, such as the matter density, the amplitude of the matter power spectrum and the equation of state of dark energy \citep[see][]{Voit2005,AllenEvrardMantz2011,KravtsovBorgani2012,Weinberg2013}. The observable properties of a galaxy cluster result from a non-trivial interplay between gravitational collapse and astrophysical processes. The diverse range of formation histories of the cluster population leads to scatter in the observable-mass scaling relations and, as surveys select clusters based on an observable, this can lead to a biased sample of clusters, resulting in systematics when using them as a cosmological probe \citep[e.g.][]{Mantz2010}. Many previous studies have shown that the relationship between a cluster observable, such as its temperature or X-ray luminosity, and a quantity of interest for cosmology, e.g. its mass, has a smaller scatter for more massive, dynamically relaxed objects \citep{EkeNavarroFrenk1998,Kay2004,Crain2007,NagaiVikhlininKravtsov2007a,Planelles2013}. Therefore, the fundamental requirement when probing cosmological parameters with galaxy clusters is a sample of relaxed, massive clusters with well calibrated mass-observable scaling relations.

However, galaxy clusters are rare objects, becoming increasingly rare with increasing mass, and to observe a sample large enough to be representative of the underlying population requires a survey with significant size and depth. Currently ongoing and impending observational campaigns, such as the Dark Energy Survey \citep{DEScol2005}, \textit{eRosita} \citep{Merloni2012}, \textit{Euclid} \citep{Laureijs2011}, \textsc{SPT-3G} \citep{Benson2014} and Advanced ACTpol \citep{Henderson2016}, will be the first to have sufficient volume to yield significant samples of massive clusters. Due to their rarity, the majority of these massive clusters will be at high redshift and it is therefore critical to understand how the cluster observables and their associated scatter evolve. Additionally, the most massive clusters will be the brightest and easiest to detect objects at high redshift, making it vital to understand the selection function of the chosen cluster observable and whether the most massive clusters are representative of the underlying cluster population. Theoretical modelling of the formation of clusters and their observable properties is required to understand these issues and to further clusters as probes of cosmology. Due to the range of scales involved in cluster formation, the need to incorporate astrophysical processes and to self-consistently predict observable properties, cosmological hydrodynamical simulations are the only viable option.

Recent progress in the modelling of large-scale structure formation has been driven mainly by the inclusion of supermassive black holes and their associated Active Galactic Nucleus (AGN) feedback, which has been shown to be critical for reproducing many cluster properties \citep{Bhattacharya2008,PuchweinSijackiSpringel2008,McCarthy2010,Fabjan2010}. A number of independent simulations are now able to produce realistic clusters that simultaneously reproduce many cluster properties in good agreement with the observations \citep{LeBrun2014,Pike2014,Planelles2014}. Results from the recent BAryons and HAloes of MAssive Systems (BAHAMAS) simulations \citep{McCarthy2016} have shown that by calibrating the subgrid model for feedback to match a small number of key observables, in this case the global galaxy stellar mass function and the gas fraction of clusters, simulations of large-scale structure are now able to reproduce many observed scaling relations and their associated scatter over two decades in halo mass. However, full gas physics simulations of large-scale structure formation, with sufficient resolution, are still computationally expensive. This has limited previous studies to either small samples with $<50$ objects or to volumes of $596\,\mathrm{Mpc}$, all of which are too small to contain the representative sample of massive clusters that is required for cosmological studies above $z=0$. 

This paper introduces the Virgo consortium's MACSIS project, a sample of $390$ massive clusters selected from a large volume dark matter simulation and resimulated with full gas physics to enable self-consistent observable predictions. The simulations extend the BAHAMAS simulations to the most massive clusters expected to form in a $\Lambda\rm{CDM}$ cosmology. In this paper we study the cluster scaling relations and their evolution. We combine the MACSIS and BAHAMAS simulations to produce a sample that spans the complete mass range and that can be studied to high redshift, using the progenitors of the MACSIS sample. We also select the hottest clusters from the combined sample and a relaxed subset of them to examine the impact of such selections on the scaling relations and their evolution. We then study the gas profiles to further understand the differences between the samples. 

This paper is organised as follows. In Section \ref{sec:MACsamp} we introduce the MACSIS sample and discuss the parent dark matter simulation from which the sample was selected, the selection criteria used, the model used to resimulate the haloes, how we produced the observable quantities and the three samples we use in this work. In Section \ref{sec:screlations} we investigate how the scaling relations evolve and how this evolution changes when a hot cluster sample or relaxed, hot cluster sample is selected. We then study the hot gas profiles to understand the differences in the evolution of the relations for the different samples in Section \ref{sec:gasprofs}. Finally, in Section \ref{sec:sad} we discuss our results and summarise our main findings.

\section{Parent simulation and sample selection}
\label{sec:MACsamp}
In this section we describe the parent simulation, the selection of the MACSIS sample, the baryonic physics used in the resimulation of the sample and the calculation of the observable properties of the resimulated clusters. Additionally, we describe how MACSIS and BAHAMAS clusters were selected to produce the combined sample and the cuts made to yield a hot sample and its relaxed subset.

\subsection{The parent simulation}
To obtain a population of massive clusters we require a simulation with a very large volume $(> 1\,\rm{Gpc}^{3})$. With current computational resources it is unfeasible to simulate such a volume with hydrodynamics and the required gas physics, such as radiative cooling, star formation and feedback, at a resolution high enough to accurately capture the cluster properties. An alternative option is to apply the zoomed simulation technique to a representative sample of objects from a larger volume. Therefore, we select a sample of massive haloes from a dark matter only simulation that has sufficient volume to yield a population of massive clusters and the resolution to ensure they are well characterized. We label this simulation the `parent' simulation.

The parent simulation is a periodic cube with a side length of $3.2\,\rm{Gpc}$. Its cosmological parameters are taken from the Planck 2013 results combined with baryonic acoustic oscillations, WMAP polarization and high multipole moments experiments \citep{Planck2014I} and are $\Omega_{\rm{b}}=0.04825$, $\Omega_{\rm{m}}=0.307$, $\Omega_{\Lambda}=0.693$, $h\equiv H_0/(100\,\rm{km}\,\rm{s}^{-1}\,\rm{Mpc}^{-1})=0.6777$, $\sigma_{8}=0.8288$, $n_{\rm{s}}=0.9611$ and $Y=0.248$. We note that there are minor differences between these values and the Planck-only cosmology used for the BAHAMAS simulations, but this has negligible impact on the results presented here. The simulation contained $N=2520^3$ dark matter particles that were arranged in an initial glass-like configuration and then displaced according to second-order Lagrangian perturbation theory $(\mathrm{2LPT})$ using the \textsc{ic\_2lpt\_gen} code \citep{Jenkins2010} and the public Gaussian white noise field \textit{Panphasia} \citep{Jenkins2013,JenkinsBooth2013}. \footnote{The phase descriptor for this volume is [Panph1, L14, (2152, 5744, 757), S3, CH1814785143, EAGLE\_L3200\_VOL1].} The particle mass of this simulation is $m_{\rm{DM}}=5.43\times10^{10}\,\mathrm{M}_{\rm{\odot}}/h$ and the comoving gravitational softening length was set to $40\,\mathrm{kpc}/h$. The simulation was evolved from redshift $z=127$ using a version of the Lagrangian TreePM-SPH code \textsc{p-gadget3} \citep[last described in][]{Springel2005}. Haloes were identified at $z=0$ using a \textit{Friends-of-Friends} (FoF) algorithm with a standard linking length of $b=0.2$ in units of the mean interparticle separation \citep{Davis1985}.

\begin{figure}
 \includegraphics[width=\columnwidth]{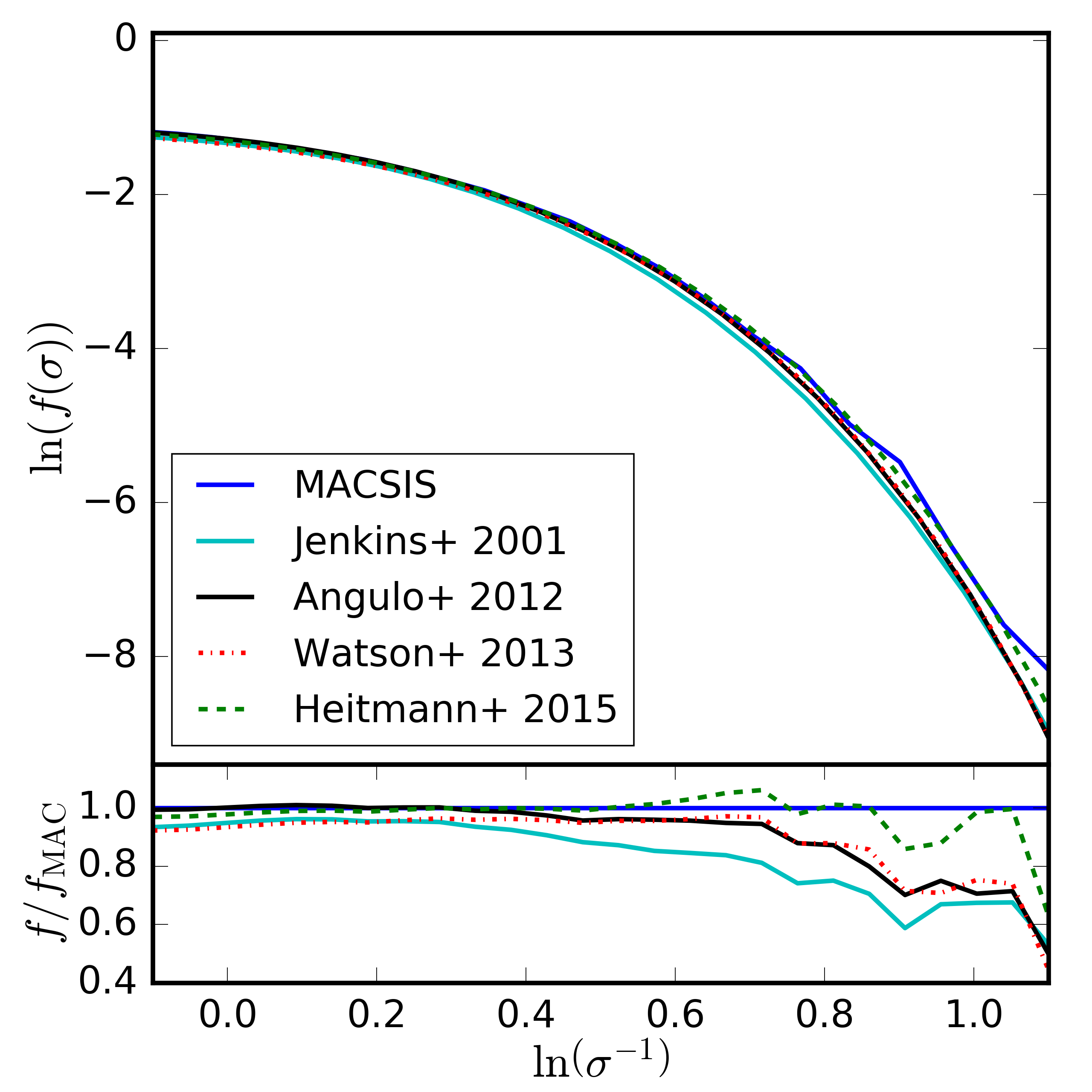}
 \caption{Comparison of the \textit{Friends-of-Friends} mass function of the parent simulation against those from \citet{Jenkins2001}, \citet{Angulo2012}, \citet{Watson2013} and \citet{Heitmann2015} (top) with the residual differences (bottom). We find good agreement with \citet{Heitmann2015}, but for values of $\ln(\sigma^{-1})>0.4$ we find a growing discrepancy between the parent simulation and the other simulations. This is likely due to our use of $2^{\mathrm{nd}}$ order Lagrangian perturbation theory when generating the initial conditions of the parent simulation and cosmic variance for the rarest haloes.}
 \label{fig:massfunc}
\end{figure}

We plot the FoF mass function of the parent simulation at $z=0$ in Fig. \ref{fig:massfunc}. We compare it to the published relations of \citet{Jenkins2001}, \citet{Angulo2012}, \citet{Watson2013} and \citet{Heitmann2015}. We plot the scaled differential mass function
\begin{equation}
    f(\sigma) = \frac{M}{\bar{\rho}}\frac{dn}{d\ln\sigma^{-1}}(M,z)\:,
\end{equation}
where $M$ is halo mass, $\bar{\rho}$ is the mean density of the Universe at $z=0$, $n$ is the number of haloes per unit volume, and $\sigma^{2}$ is the variance of the linear density field when smoothed with a top-hat filter. We plot the mass function as a function of the variable $\ln(\sigma^{-1})$ as it is insensitive to cosmology \citep{Jenkins2001}. For $\ln(\sigma^{-1}) < 0.3$ we find that all of the mass functions show reasonable agreement with differences of $\sim5-10\%$ between them, with the small differences likely due to the mass function not being exactly universal \citep{Tinker2008,Courtin2011}. However, for larger values the mass functions begin to diverge, as the parent simulation has an excess of massive clusters compared to the other simulations. This is likely due to two effects. First, the MACSIS simulation is the only one to use $\mathrm{2LPT}$ when generating the initial conditions. It has been shown that not using $\mathrm{2LPT}$ results in a significant underestimation of the abundance of the rarest objects \citep{Crocce2006,Reed2013}. The second effect is simply statistics: even in a very large volume there are still low numbers of the rarest and most massive clusters, where there is likely to be significant variance between the simulation volumes.

\subsection{The MACSIS sample}
To select the MACSIS sample, all haloes with $M_{\rm{FoF}}>10^{15}\,\mathrm{M}_{\odot}$ were grouped in logarithmically spaced bins , with $\Delta\log_{10}\,M_{\rm{FoF}}=0.2$. If a bin contained less than one hundred haloes then all of the objects in that bin were selected. For bins with more than one hundred objects the bin was then further subdivided into bins of 0.02 dex and ten objects from each sub-bin were then selected at random. The subdividing of the bins ensured that our random selection was not biased to low masses by the steep slope of the mass function. This selection procedure results in a sample of $390$ haloes that is mass limited above $10^{15.6}\,\rm{M}_{\odot}$ and randomly sampled below this limit. Table \ref{tab:smpcomp} shows the fraction of haloes selected from the parent simulation in each mass bin. We have compared the properties of the selected haloes with those of the underlying population and found the MACSIS sample to be representative. Additionally, in Appendix \ref{app:seleff} we demonstrate that selecting by a halo's FoF mass does not bias our results when binning clusters by their $M_{500}$.

\begin{table}
 \caption{Table showing the fraction of haloes from the parent simulation that are part of the MACSIS sample for the selection mass bins. The sample is complete above $M_{\rm{FoF}} > 10^{15.6}\,\mathrm{M}_{\odot}$. The parent simulation contains $9754$ haloes with $M_{\rm{FoF}} > 10^{15.0}\,\mathrm{M}_{\odot}$ at $z=0$.}
 \centering
 \begin{tabular}{l r r c}
  \hline
  Mass Bin & \multicolumn{1}{c}{Sample} & \multicolumn{1}{c}{Total} & Fraction \\
   & \multicolumn{1}{c}{Size} & \multicolumn{1}{c}{Haloes} & Selected \\
  \hline
  $15.0 \leq \log_{10}(M_{\rm{FoF}}) < 15.2$ & $100~~$ & $7084~$ & $0.01$ \\
  $15.2 \leq \log_{10}(M_{\rm{FoF}}) < 15.4$ & $100~~$ & $2095~$ & $0.05$ \\
  $15.4 \leq \log_{10}(M_{\rm{FoF}}) < 15.6$ & $100~~$ & $485~$ & $0.21$ \\
  $15.6 \leq \log_{10}(M_{\rm{FoF}}) < 15.8$ & $83~~$ & $83~$ & $1.00$ \\
  $15.8 \leq \log_{10}(M_{\rm{FoF}})$ & $7~~$ & $7~$ & $1.00$ \\
  \hline
 \end{tabular}
 \label{tab:smpcomp}
\end{table}

\begin{figure*}
 \includegraphics[width=\textwidth]{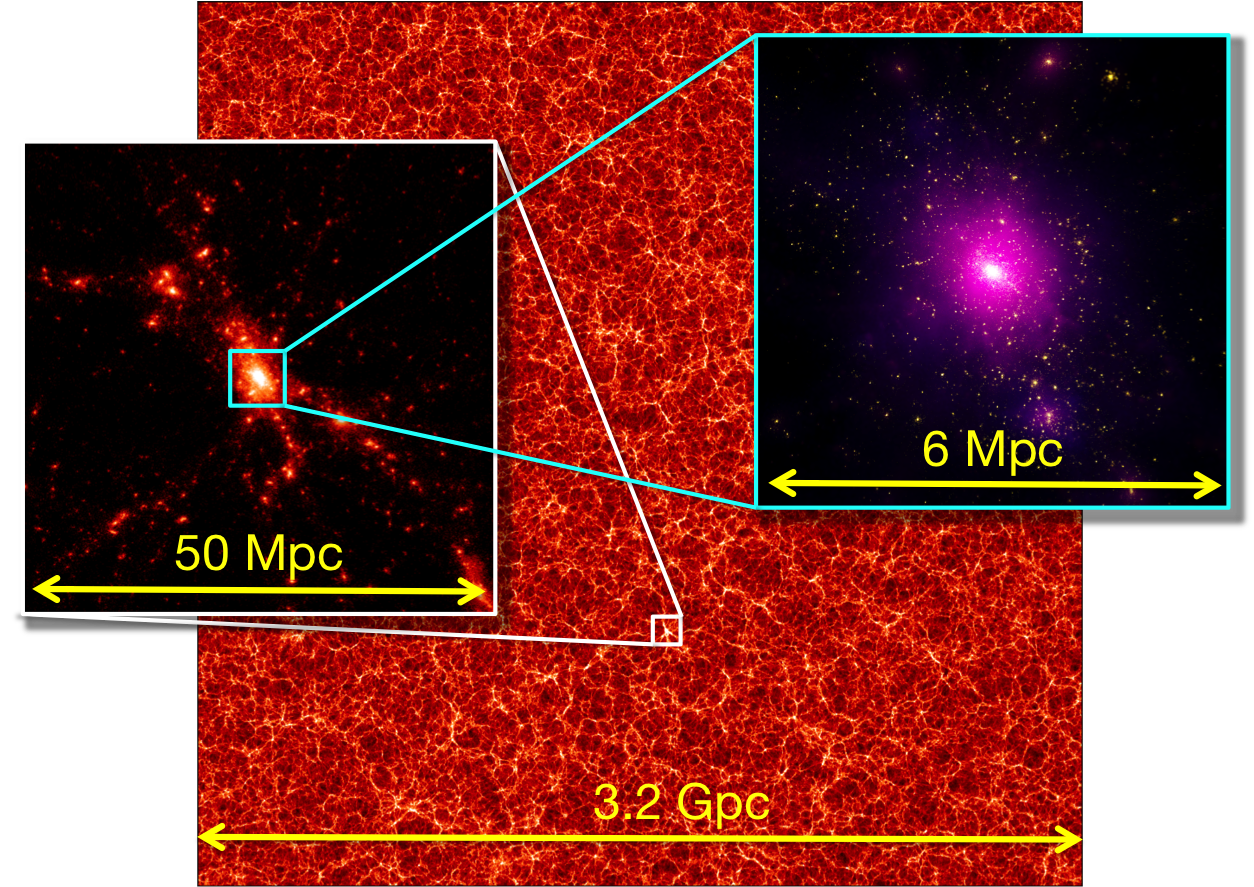}
 \caption{Slice of depth $40\,\rm{Mpc}$ through the parent simulation showing the projected dark matter density at $z=0$. The left inset shows a $50\,\rm{Mpc}$ cube centred on the most massive halo. The right inset shows the stellar particles of the same halo in yellow, re-simulated using the BAHAMAS model and resolution, with X-ray emission from the hot gas overlaid in purple.}
 \label{fig:zoomed}
\end{figure*}

Due to current computational constraints the BAHAMAS simulations are limited to periodic cubes with a side length of $596\,\mathrm{Mpc}$. There are very few clusters with a mass greater than $10^{15}\,\mathrm{M}_{\odot}$ in a volume of this size, and those that are present may be affected by the loss of power from large-scale modes that are absent due to their wavelengths being greater than the box size. The zoom simulations of the MACSIS project provide an extension to the BAHAMAS periodic simulations. They provide the most massive clusters and allow the mass-observable scaling relations to be studied across the complete cluster mass range.

We use the zoomed simulation technique \citep{KatzWhite1993,Tormen1997} to re-simulate the chosen sample at increased resolution. We perform both DM only and full gas physics re-simulations. The Lagrangian region for every cluster was selected so that its volume was devoid of lower resolution particles beyond a cluster centric radius of $5r_{200}$.\footnote{We define $r_{200}$ as the radius at which the enclosed average density is two hundred times the critical density of the Universe.} The resolution of the Lagrangian region was increased such that the particles in the DM only simulations had a mass of $m_{\rm{DM}}=5.2\times10^9\,\mathrm{M}_{\odot}/h$ and in the hydrodynamic re-simulations the dark matter particles had a mass of $m_{\rm{DM}}=4.4\times10^9\,\mathrm{M}_{\odot}/h$ and the gas particles had an initial mass of $m_{\rm{gas}}=8.0\times10^8\,\mathrm{M}_{\odot}/h$. In all simulations the Plummer equivalent gravitational softening length for the high-resolution particles was fixed to $4\mathrm{kpc}/h$ in comoving units for $z>3$ and in physical coordinates thereafter. The smoothed particle hydrodynamics interpolation used $48$ neighbours and the minimum smoothing length was set to one tenth of the gravitational softening. A schematic view of the zoom approach is shown in Fig. \ref{fig:zoomed}.

The resolution and softening of the zoom re-simulations were deliberately chosen to match the values of the periodic box simulations of the BAHAMAS project \citep{McCarthy2016}, which is a calibrated version of the OWLS code \citep{Schaye2010}, which was also used for cosmo-OWLS \citep{LeBrun2014}. The subgrid models for feedback from star formation and AGN used in the BAHAMAS simulations was calibrated to obtain a good fit to the observed galaxy stellar mass function and the amplitude of the gas fraction-total mass relation, respectively, at $z=0$. Without any further tuning, the simulations then produce a population of groups and clusters that shows excellent agreement with the observations for a range of galaxy-halo, hot gas-halo and galaxy-hot gas relations.

\subsection{Baryonic physics}
The BAHAMAS simulations were run with a version of \textsc{p-gadget3} that has been heavily modified to include new subgrid physics as part of the OWLS project \citep{Schaye2010}. We now briefly describe the subgrid physics, but refer the reader to \citet{Schaye2010}, \citet{LeBrun2014} and \citet{McCarthy2016} for greater detail, including the impact of varying the free parameters in the model and the calibration strategy. Radiative cooling is calculated on an element-by-element basis following \citet{WiersmaSchayeSmith2009}, interpolating the rates as a function of density, temperature and redshift from pre-computed tables generated with \textsc{cloudy} \citep{Ferland1998}. It accounts for heating and cooling due to the primary cosmic microwave background and a \citet{HaardtMadau2001} ultra-violet/X-ray background. The background due to reionization is assumed to switch on at $z=9$.

Star formation is modelled stochastically in a way that by construction reproduces the observations, as discussed in \citet{SchayeDallaVecchia2008}. Lacking the resolution and physics to correctly model the cold interstellar medium, gas particles with a density $n_{\rm{H}} > 0.1\,\rm{cm}^{-3}$ follow an imposed equation of state with $P\propto\rho^{4/3}$. These gas particles then form stars at a pressure-dependent rate that reproduces the observed Kennicutt-Schmidt law \citep{Schmidt1959,Kennicutt1998}. Stellar evolution and the resulting chemical enrichment are implemented using the model of \citet{Wiersma2009}, where $11$ chemical elements (H, He, C, N, O, Ne, Mg, Si, S, Ca and Fe) are followed. The mass loss rates are calculated assuming Type Ia and Type II supernovae and winds from massive and asymptotic giant branch stars. Stellar feedback is implemented via the kinetic wind model of \citet{DallaVecchiaSchaye2008}. The BAHAMAS simulations used the calibrated mass-loading factor of $\eta_{\rm{w}}=2$ and wind velocity $v_{\rm{w}}=300\,\rm{km/s}$. This corresponds to $20$ percent of available energy from Type II supernovae, assuming a \citet{Chabrier2003} IMF, and yields an excellent fit to the observed galaxy mass function.

The seeding, growth and feedback from supermassive black holes (BH) is implemented using the prescription of \citet{BoothSchaye2009}, a modified version of the method developed by \citet{SpringelDiMatteoHernquist2005}. A FoF algorithm is run on-the-fly and BH seed particles, with $m_{\mathrm{BH}}=10^{-3}m_{\mathrm{gas}}$, are placed in haloes that contain at least $100$ DM particles, which corresponds to a halo mass of $\sim5\times10^{11}\,\rm{M}_{\odot}$. BHs grow via Eddington-limited accretion of gas at the Bondi-Hoyle-Littleton rate, with a boost factor that is a power-law of the local density for gas above the star formation density threshold. They also grow by direct mergers with other BHs. A fraction, $\epsilon$, of the rest mass energy of the accreted gas is then used to heat $n_{\rm{heat}}$ neighbour particles by increasing their temperature by $\Delta T_{\rm{heat}}$. Changes to these parameters have a significant impact on the hot gas properties of clusters. The calibrated values of these parameters in the BAHAMAS simulations are $n_{\rm{heat}}=20$ and $\Delta T_{\rm{heat}}=10^{7.8}\,\rm{K}$. The feedback efficiency $\epsilon=\epsilon_{\rm{r}}\epsilon_{\rm{f}}$, where $\epsilon_{\rm{r}}=0.1$ is the radiative efficiency and $\epsilon_{\rm{f}}=0.15$ is the fraction of $\epsilon_{\rm{r}}$ that couples to the surrounding gas. The choice of the efficiency, assuming it is non-zero, is generally of little consequence as the feedback establishes a self-regulating scenario, but determines the black hole masses \citep{BoothSchaye2009}. 

\subsection{Calculating observable properties}
Previous studies have shown that there can be significant biases in the observable properties of clusters due to issues such as multi-temperature structures and gas inhomogeneities \citep[e.g.][]{NagaiVikhlininKravtsov2007b,Khedekar2013}. Therefore, when investigating cluster properties it is critical that, as far as possible, we make a like-with-like comparison with the observations. Following \citet{LeBrun2014}, we do this by producing synthetic observational data for each cluster and analysing it in a manner similar to what is done for real data. Using the particle's temperature, density and metallicity, where the metallicity is smoothed over a particle's neighbours, we first compute a rest-frame X-ray spectrum in the $0.05-100.0\,\rm{keV}$ band for all gas particles, using the Astrophysical Plasma Emission Code \citep[\textsc{apec};][]{Smith2001} via the \textsc{pyatomdb} module with atomic data from \textsc{atomdb} v3.0.2 \citep[last described in][]{Foster2012}. A particle's spectrum is a sum of the individual spectra for each chemical element tracked by the simulations, scaled by the particle's elemental abundance. We ignore particles with a temperature lower than $10^5\,\rm{K}$ as they make a negligible contribution to the total X-ray emission.

We then estimate the density, temperature and metallicity of the hot gas in $25$ logarithmically spaced radial bins by fitting a single temperature \textsc{apec} model, with a fixed metallicity, to the summed spectra of all particles that fall within that radial bin. We then scale the spectra by the relative abundance of the heavy elements as the fiducial spectra assume solar abundance \citep{AndersGrevesse1989}. The spectra have an energy resolution of $150\,\rm{eV}$ in the range $0.05-10.0\,\rm{keV}$ and are logarithmically spaced between $10.0-100.0\,\rm{keV}$. To get a closer match to the observations, we multiply the spectra by the effective area of \textit{Chandra}. To derive temperature and density profiles of a cluster, we fit the spectrum in the range $0.5-10.0\,\rm{keV}$ for each radial bin with a single temperature model using a least-squares approach. 

The temperature and density profiles derived from the X-ray spectra are then used to perform a hydrostatic mass analysis of the cluster. The profiles are fit with the density and temperature models proposed by \citet{Vikhlinin2006} to produce a hydrostatic mass profile. We then derive various mass and radius estimates, such as $M_{500}$ and $r_{500}$, from the hydrostatic mass profiles. With these estimates we calculate quantities, such as $M_{\rm{gas}}$ or $Y_{\rm{SZ}}$, by summing the properties of the particles that fall within the set. Core-excised quantities are calculated in the radial range $0.15-1.0$ of the aperture. Luminosities are calculated by integrating the spectra of all particles within the aperture in the requisite energy band, for example, bolometric luminosities are calculated in the range $0.05-100.0\,\rm{keV}$. Averaged X-ray temperatures are calculated by fitting a single temperature model to the sum of the spectra of all particles within the aperture. We repeat this analysis for all clusters in the combined sample at all redshifts of interest. All quantities derived in this manner are labelled with the sub-script `spec'.

\subsection{Cluster sample selection}
\begin{figure*}
  \includegraphics[width=\textwidth]{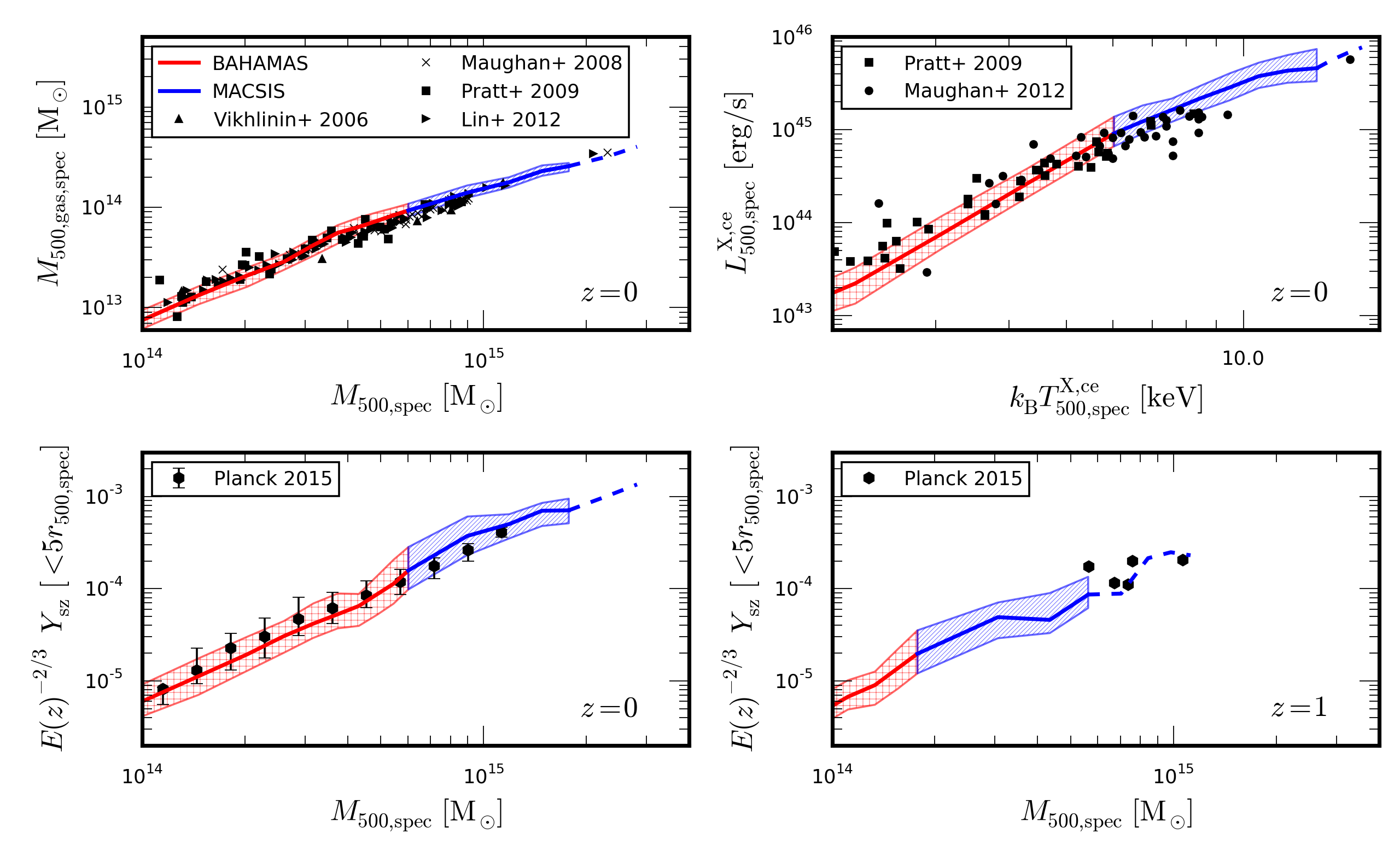}
  \caption{Gas mass-total mass relation (top left), core-excised bolometric X-ray luminosity-core-excised X-ray temperature relation (top right) and the integrated Sunyaev-Zel'dovich signal-total mass relation at $z=0$ (bottom left) and $z=1$ (bottom right) for the combined sample. The median relation of the BAHAMAS sample is given by the red line, with the red hatch region enclosing $68\%$ of the population, and the median MACSIS result is shown by the blue line, with the blue hatched region enclosing $68\%$ of the sample. The median MACSIS line becomes dashed when there are less than $10$ clusters in a bin. The black triangles, crosses, squares, right-facing triangles, circles, left-facing triangles, hexagons and pluses are observational data from \citet{Vikhlinin2006}, \citet{Maughan2008}, \citet{Pratt2009}, \citet{Lin2012}, \citet{Maughan2012} and the second \textit{Planck} SZ catalogue \citep{PlanckSZ2015} respectively.}
  \label{fig:observations}
\end{figure*}

We select clusters from MACSIS and BAHAMAS to form a `combined' sample with which we can investigate the cluster scaling relations. We perform our analysis at $z=0.0,0.25,0.5,1.0$ and $1.5$. We create this sample at each redshift by selecting all clusters with a mass $M_{500,\rm{spec}}\geq10^{14}\,\rm{M}_{\odot}$. Additionally, we introduce a mass cut at every redshift below which we remove any MACSIS clusters. For example, at $z=0$ ($z=1$) this cut is made at $M_{500,\mathrm{spec}}=10^{14.78}\,\rm{M}_{\odot}$ ($M_{500,\mathrm{spec}}=10^{14.3}\,\rm{M}_{\odot}$). This removes a tail of clusters with low $M_{500,\mathrm{spec}}$, but have high $M_{\mathrm{FoF}}/M_{500,\mathrm{spec}}$ ratios (see Appendix \ref{app:seleff}). For the luminosity-temperature relation, we use the temperature-mass relation of the combined sample to convert the mass cut into a temperature cut. At $z=0$ this results in a sample of $1294$ clusters, containing $1098$ clusters from BAHAMAS and $196$ MACSIS clusters, and at $z=1$ a sample of $225$ clusters, $99$ from BAHAMAS and $126$ from MACSIS.

The MACSIS clusters enable the investigation of the behaviour of the most massive clusters at low redshift. These clusters are commonly selected in cosmological analyses because their deep potentials are expected to reduce the impact of non-gravitational processes and as the brightest clusters they require shorter exposures. We select a hot, and therefore massive, cluster sample by selecting all clusters in the combined sample with a core-excised X-ray temperature greater than $5\,\rm{keV}$. At $z=0$ ($z=0.5$) this yields a sample of $244$ ($186$) clusters, with $190$ ($173$) coming from the MACSIS sample. Finally, we examine the impact of selecting a relaxed subset of the hot cluster sample. Theoretically, there are many ways to define a relaxed halo \citep[see][]{Neto2007,Duffy2008,Klypin2011,DuttonMaccio2014,Klypin2016}. For this study we use the following criteria
\begin{equation}
 X_{\rm{off}} < 0.07\,;~f_{\rm{sub}} < 0.1~\rm{and}~\lambda < 0.07\,, \nonumber
\end{equation}
where $X_{\rm{off}}$ is distance between the cluster's minimum gravitational potential and centre of mass, divided by its virial radius; $f_{\rm{sub}}$ is the mass fraction within the virial radius that is bound to substructures; and $\lambda$ is the spin parameter for all particles inside $r_{200}$. These criteria are not designed to select a small subset that comprises the most relaxed objects, but to simply remove those clusters that are significantly disturbed. This results in a subsample at $z=0$ ($z=0.5$) that contains $213$ ($117$) clusters, with $177$ ($111$) coming from the MACSIS sample.

\section{The scaling relations of massive clusters}
\label{sec:screlations}

In this section we present our main results, measuring the scaling relations of our cluster samples across a range of redshifts.

\subsection{Comparison to observational data}
Fig. \ref{fig:observations} shows the gas mass, $M_{\rm{gas},500,\rm{spec}}$, the integrated Sunyaev-Zel'dovich (SZ) signal, $Y_{\rm{SZ}}$, measured in a $5r_{500,spec}$ aperture as a function of estimated total mass, $M_{500,\rm{spec}}$, (at $z=0$ and $z=1$) and the core-excised bolometric X-ray luminosity, $L^{\rm{X,ce}}_{500,\rm{spec}}$, as a function of core-excised X-ray temperature, $T^{\rm{X,ce}}_{500,\rm{spec}}$, for the combined sample. We compare the sample to the relevant observational data. At all redshifts the MACSIS sample provides a consistent extension to the BAHAMAS clusters with similar scatter. At low redshift, \citet{McCarthy2016} have shown that the BAHAMAS sample shows good agreement with the observed median relations and shows similar intrinsic scatter. The MACSIS sample continues this agreement to observed high-mass clusters, though there are significantly fewer clusters to compare against. In detail, it appears that the $M_{500,\mathrm{gas,spec}}$-$M_{500,\mathrm{spec}}$ and $L^{\mathrm{X,ce}}_{500,\mathrm{spec}}$-$T^{\mathrm{X,ce}}_{500,\mathrm{spec}}$ relations are slightly steeper than observed. However, we would exercise caution as we have not applied the same selection criteria as was used for the observational X-ray analyses.

At high redshift observational data becomes sparse and currently only SZ surveys have detected a reasonable number of clusters. At $z=1$ these clusters are all significantly more massive than any cluster in the BAHAMAS volume. However, the progenitors of the very massive MACSIS clusters provide a sample that can be compared with these observations. We find that the median relation shows good agreement with the observations and the intrinsic scatter of the clusters about the median relation is consistent with the scatter in the observations. Overall, we find that all quantities computed in a like-with-like manner show good agreement with the observations.

\renewcommand\arraystretch{1.5}
\newcolumntype{C}{>{\centering\arraybackslash}X}
\begin{table*}
 \caption{The normalization and slope of the best-fit relations presented in this work and the scatter about them for the three samples at $z=0$. All quantities presented in this table are `$\rm{spec}$' values calculated via the synthetic X-ray analysis within an aperture of $r_{500,\rm{spec}}$. The scatter $\langle\sigma_{\log_{10}Y}\rangle$ is averaged over all masses.}
 \centering
 \begin{tabularx}{\textwidth}{l r C C r C C r C C}
  \hline
  Scaling relation & \multicolumn{3}{c}{Combined sample} & \multicolumn{3}{c}{Hot Clusters} & \multicolumn{3}{c}{Relaxed, Hot Clusters} \\
   & \multicolumn{1}{c}{$A$} & \multicolumn{1}{c}{$\alpha$} & \multicolumn{1}{c}{$\langle\sigma_{\log_{10}Y}\rangle$} & \multicolumn{1}{c}{$A$} & \multicolumn{1}{c}{$\alpha$} & \multicolumn{1}{c}{$\langle\sigma_{\log_{10}Y}\rangle$} & \multicolumn{1}{c}{$A$} & \multicolumn{1}{c}{$\alpha$} & \multicolumn{1}{c}{$\langle\sigma_{\log_{10}Y}\rangle$} \\
  \hline
  $L^{\rm{X,ce}}_{500}-M_{500}$ & $44.50^{+0.01}_{-0.01}$ & $1.88^{+0.03}_{-0.05}$ & $0.15^{+0.01}_{-0.02}$ & $44.71^{+0.02}_{-0.02}$ & $1.36^{+0.08}_{-0.07}$ & $0.12^{+0.01}_{-0.02}$ & $44.69^{+0.03}_{-0.03}$ & $1.43^{+0.13}_{-0.09}$ & $0.11^{+0.01}_{-0.01}$ \\
  $k_{\rm{B}}T^{\rm{X,ce}}_{500}-M_{500}$ & $0.68^{+0.01}_{-0.01}$ & $0.58^{+0.01}_{-0.01}$ & $0.048^{+0.003}_{-0.003}$ & $0.71^{+0.01}_{-0.01}$ & $0.51^{+0.04}_{-0.04}$ & $0.05^{+0.01}_{-0.01}$ & $0.70^{+0.01}_{-0.01}$ & $0.55^{+0.06}_{-0.03}$ & $0.04^{+0.01}_{-0.01}$ \\
  $M_{\rm{gas},500}-M_{500}$ & $13.67^{+0.01}_{-0.01}$ & $1.25^{+0.01}_{-0.03}$ & $0.07^{+0.01}_{-0.01}$ & $13.77^{+0.01}_{-0.01}$ & $1.02^{+0.03}_{-0.03}$ & $0.06^{+0.01}_{-0.01}$ & $13.75^{+0.01}_{-0.01}$ & $1.05^{+0.04}_{-0.04}$ & $0.05^{+0.01}_{-0.01}$ \\
  $Y_{\rm{X},500}-M_{500}$ & $14.33^{+0.01}_{-0.01}$ & $1.84^{+0.02}_{-0.05}$ & $0.12^{+0.01}_{-0.01}$ & $14.47^{+0.02}_{-0.02}$ & $1.51^{+0.07}_{-0.08}$ & $0.11^{+0.01}_{-0.01}$ & $14.45^{+0.02}_{-0.02}$ & $1.59^{+0.12}_{-0.06}$ & $0.08^{+0.01}_{-0.01}$ \\
  $Y_{\rm{SZ},500}-M_{500}$ & $-4.51^{+0.01}_{-0.01}$ & $1.88^{+0.02}_{-0.03}$ & $0.10^{+0.01}_{-0.01}$ & $-4.39^{+0.02}_{-0.02}$ & $1.60^{+0.07}_{-0.05}$ & $0.10^{+0.01}_{-0.02}$ & $-4.42^{+0.02}_{-0.02}$ & $1.69^{+0.07}_{-0.07}$ & $0.09^{+0.01}_{-0.01}$ \\
  $L^{\rm{X,ce}}_{500}-T^{\rm{X,ce}}_{500}$ & $44.80^{+0.02}_{-0.01}$ & $3.01^{+0.04}_{-0.04}$ & $0.14^{+0.01}_{-0.01}$ & $44.93^{+0.01}_{-0.01}$ & $2.41^{+0.12}_{-0.12}$ & $0.11^{+0.01}_{-0.01}$ & $44.89^{+0.02}_{-0.02}$ & $2.53^{+0.12}_{-0.13}$ & $0.10^{+0.01}_{-0.01}$ \\
  \hline
 \end{tabularx}
 \label{tab:z0bestfit}
\end{table*}
\renewcommand\arraystretch{1.0}

\subsection{Modelling cluster scaling relations}
As a baseline for understanding how the scaling relations evolve as a function of mass and redshift, we adopt the following self-similar scalings
\begin{equation}
    M_{\rm{gas},\Delta}\propto M_{\Delta}\,,
\end{equation}
\begin{equation}
    \label{eq:T-M}
    T_{\Delta}\propto M^{2/3}_{\Delta}E^{2/3}(z)\,,
\end{equation}
\begin{equation}
    Y_{\rm{X},\Delta}\propto M_{\Delta}^{5/3}E^{2/3}(z)\,,
\end{equation}
\begin{equation}
    Y_{\rm{SZ},\Delta}\propto M_{\Delta}^{5/3}E^{2/3}(z)\,,
\end{equation}
\begin{equation}
    \label{eq:L-M}
    L_{\Delta}^{\rm{X,bol}}\propto M^{4/3}_{\Delta}E^{7/3}(z)\,,
\end{equation}
\begin{equation}
    L_{\Delta}^{\rm{X,bol}}\propto T^{2}E(z)\,,
\end{equation}
where $E(z)\equiv H(z)/H_0=\sqrt{\Omega_{m}(1+z)^3+\Omega_{\Lambda}}$, $\Delta$ is the chosen overdensity relative to the critical density and $Y_{\rm{X}}$ is the X-ray analogue of the integrated SZ effect. These are derived in Appendix \ref{app:ssr}. Although shown to be too simplistic by the first X-ray studies of clusters \citep{Mushotzky1984,EdgeStewart1991,David1993}, the self-similar relations allow us to investigate if astrophysical processes are less significant in more massive clusters or at higher redshift. To enable a comparison with the self-similar predictions, and previous work, we fit the scaling relations of our samples at each redshift. We derive a median relation by first binning the clusters into bins of log mass (width $0.1$ dex) or log temperature (width $0.07$ dex) and then computing the median in each bin with more than ten clusters. We also remove the evolution in normalization predicted by self-similar relations. The medians of the bins are then fit with a power-law of the form
\begin{equation}
 E^{\beta}(z)Y=10^A\left(\frac{X}{X_0}\right)^{\alpha}\,,
 \label{eq:plfit}
\end{equation}
where $A$ and $\alpha$ describe the normalization and slope of the best fit respectively, $\beta$ removes the expected self-similar evolution with redshift, $X$ is either the total mass or temperature and $Y$ is the observable quantity ($M_{\rm{gas}}$, $L^{\rm{X,bol}}$, etc.). $X_0$ is the pivot point, which we set to $4\times10^{14}\,\rm{M}_{\odot}$ for observable-mass relations and to $6\,\rm{keV}$ for observable-temperature relations. We note that we fix the pivot for all samples and all redshifts. Fitting to the medians of bins, rather than individual clusters, prevents the fit from being dominated by low-mass objects, which are significantly more abundant due to the shape of the mass function. For the hot sample and its relaxed subset there are too few bins with ten or more clusters to reliably derive a best-fit relation at $z\geq1$. By limiting our sample to systems with $M_{500}\geq10^{14}\,\rm{M}_{\odot}$ we avoid any breaks in the powerlaw relations that have been seen both observationally and in previous simulation work \citep{LeBrun2016}.

We compute the scatter about the best-fit relation at each redshift by calculating the root mean squared (\textit{rms}) dispersion in each bin according to
\begin{equation}
    \sigma_{\log_{10}Y}=\sqrt{\frac{1}{N}\sum_{i=1}^N\left[\log_{10}(Y_i)-\log_{10}(Y_{\rm{BF}})\right]^2}\,,
\end{equation}
where $i$ runs over all clusters in the bin, $Y_{\rm{BF}}$ is the best fit relation for a cluster with a value $X_i$ and we note that $\sigma_{\ln Y}=\ln(10)\sigma_{\log_{10}Y}$. We obtain the uncertainties for our fit parameters by bootstrap re-sampling the clusters $10,000$ times. The best-fit values of all the scaling relations considered for the three samples (combined, hot and relaxed) at $z=0$ are summarized in Table \ref{tab:z0bestfit} and other redshifts are listed in Appendix \ref{app:fitpar}. We now discuss each relation in turn.

\subsection{Gas Mass-Total Mass}
\label{ssec:Mg-M}
\begin{figure*}
 \includegraphics[width=\textwidth]{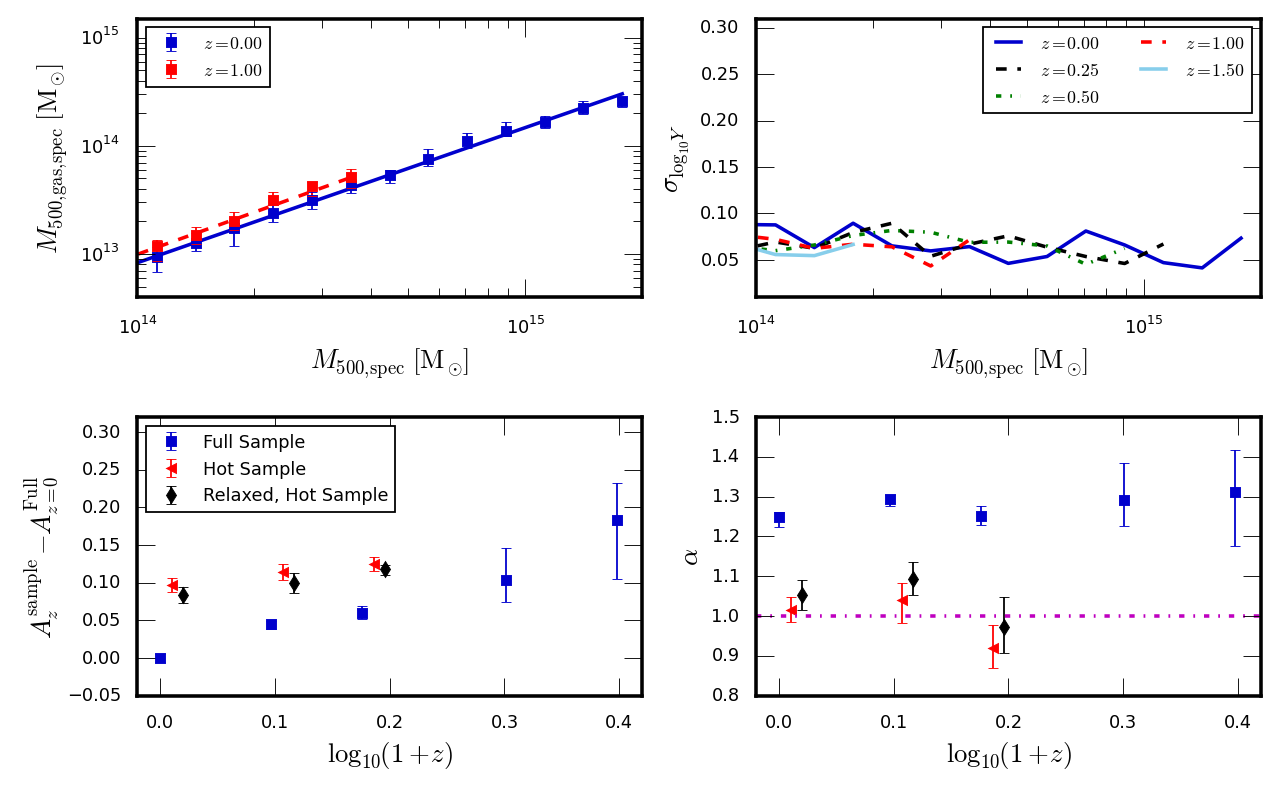}
 \caption{Evolution of the gas mass-total mass scaling relation for the three samples as a function of redshift. The top left panel shows the median gas mass in bins of total mass at $z=0$ (blue) and $z=1$ (red) for the combined sample, with error bars showing the $16^{\rm{th}}$ and $84^{\rm{th}}$ percentiles of the distribution in each bin. The solid (dashed) line shows the best-fit relation at $z=0$ ($z=1$). Note that only two redshifts are shown for clarity. The top right panel shows the \textit{rms} scatter in each mass bin at each redshift for the combined sample. The bottom panels show the best-fit normalization, $A$, (left) and slope, $\alpha$, (right) of the scaling relation as a function of $\log_{10}(1+z)$ for the three different samples: combined (blue squares), hot clusters (red triangles) and relaxed hot sample (black diamonds). We have offset the points for clarity. The dot-dashed magenta line shows the value of the predicted self-similar slope.}
 \label{fig:MgMsr}
\end{figure*}

We plot the hot gas mass-total mass scaling relation for the three samples in Fig. \ref{fig:MgMsr}. The best-fit normalization for the combined sample shows significant evolution with redshift, with clusters of a fixed mass containing $25\%$ more hot gas at $z=1$ than at $z=0$. With the inclusion of star formation, radiative cooling and feedback from supernovae and AGN, the departure from self-similarity is not unexpected. The increasing normalization with redshift is due to either the impact of AGN feedback or the conversion of gas to stars. As the normalization of the baryonic mass exhibits a similar trend, this evolution is being driven by AGN feedback. A plausible explanation is as follows. The mean density of the Universe increases with redshift and cluster potentials at a fixed mass get deeper with increasing redshift. This reduces the efficiency with which AGN expel gas from the cluster with increasing redshift, leading to a higher gas mass at higher redshift for clusters at a fixed mass. In addition, AGN have less time to act on and expel gas from clusters that form at higher redshifts. The AGN breaks the self-similar assumption of a constant gas fraction, resulting in the normalization of the gas mass-total mass relation increasing with increasing redshift. However, we note that this behaviour appears to be dependent on the implementation of the subgrid physics. \citet{LeBrun2016} use the same subgrid implementation, but with different parameters, and obtain similar behaviour. However, \citet{Planelles2013} see a constant baryon fraction with redshift suggesting that feedback is not expelling gas beyond $r_{500}$. 

The bottom left panel of Fig. \ref{fig:MgMsr} shows that the normalizations of the best-fit relations for the hot sample of clusters and for the relaxed subset of hot clusters are higher at $z=0$ than the normalization of the combined sample and evolve less with redshift. This is because hotter clusters are generally more massive and have deeper potential wells, reducing the amount of gas the AGN can permanently expel from the cluster during its formation. This flattens the slope of the relation leading to a higher normalization at the pivot.

The bottom right panel of Fig. \ref{fig:MgMsr} shows that the slope of the best-fit relation of the combined sample is significantly steeper than the self-similar prediction of unity. At a given redshift AGN feedback has expelled more gas from lower mass clusters, due to their shallower potentials, leading to a tilt in the relation. We find a slope of $\alpha=1.25^{+0.01}_{-0.03}$. Our slope is mildly shallower than found in previous a simulation work, where \citet{LeBrun2016} find a slope of $1.32$ for their AGN8.0 simulation, but consistent with observations, where \citet{Arnaud2007} found a slope of $1.25\pm0.06$ for a sample of clusters observed with \textit{XMM}. We find negligible evolution in the slope of the relation for the combined sample.

The hot cluster sample and the relaxed subset have best-fit slopes that are consistent with the self-similar prediction. The increased depth of the potential well in massive clusters means that their gas mass is approximately a constant fraction of their total mass. Specifically, we find that most massive clusters have a median gas fraction $f_{\mathrm{gas}}=0.89\pm0.09$ of the universal baryon fraction at $z=0$. This results in slopes of $\alpha=1.02\pm0.03$ and $1.05\pm0.04$ for the hot cluster sample and the relaxed subset respectively. We find good agreement with the slope of $1.05\pm0.05$ found by \citet{Mantz2016} and the self-similar slope found by \citet{Vikhlinin2009} for relaxed cluster samples. The slope of the best-fit relation for both samples shows no significant evolution with redshift.

The top right panel of Fig. \ref{fig:MgMsr} shows that the scatter about the best-fit relation is independent of both mass and redshift. Averaged over all mass bins it has a value of $\sigma_{\log_{10}Y}=0.07$ at $z=0$. The scatter reduces slightly for the hot cluster sample, with a value of $0.06$, and further still for the relaxed subset, with a value of $0.05$. The scatter is in reasonable agreement with the scatter of $0.04$ found by \citet{Arnaud2007} for a sample of clusters observed with \textit{XMM}.

\begin{figure*}
 \includegraphics[width=\textwidth]{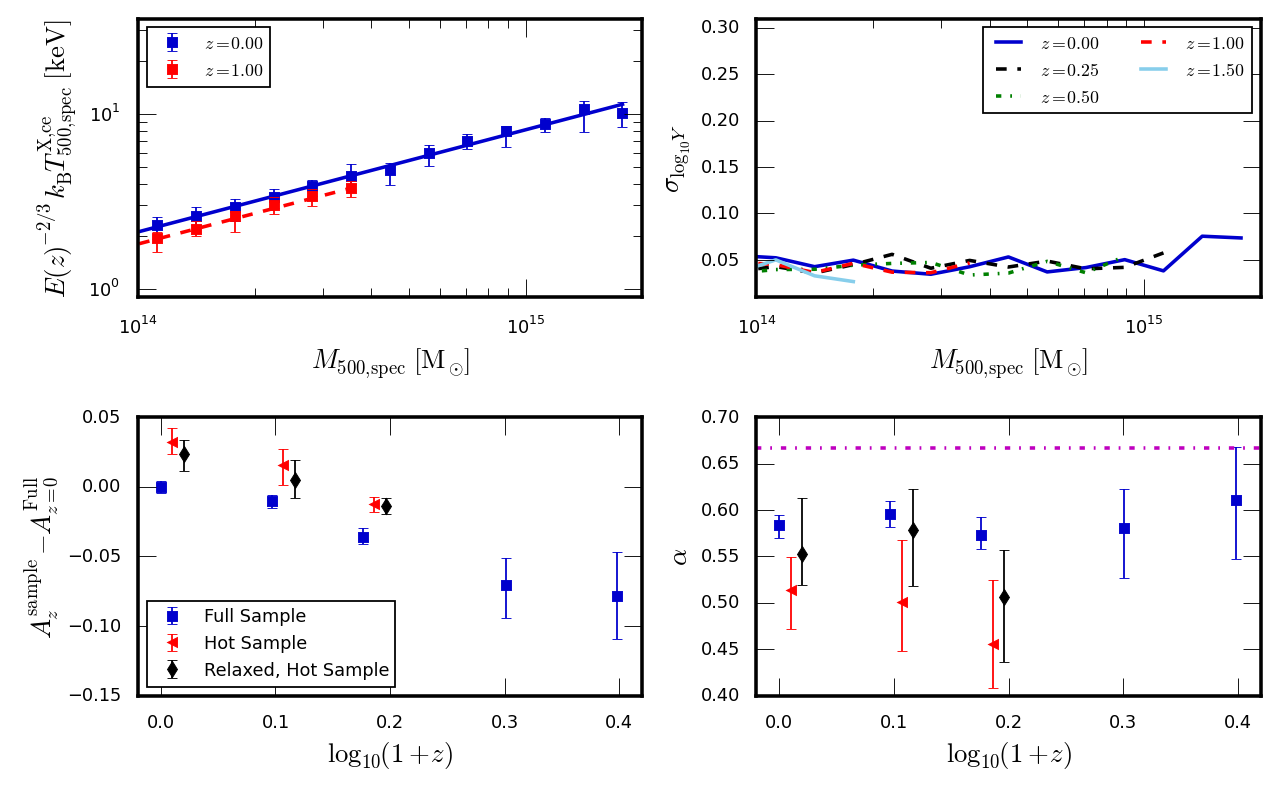}
 \caption{Evolution of the core-excised X-ray temperature-total mass scaling relation for the three samples as a function of redshift. The panels are arranged as described in Fig. \ref{fig:MgMsr}.}
 \label{fig:TxMsr}
\end{figure*}

\subsection{X-ray Temperature-Mass}
\label{ssec:Tx-M}
\begin{figure}
 \includegraphics[width=\columnwidth]{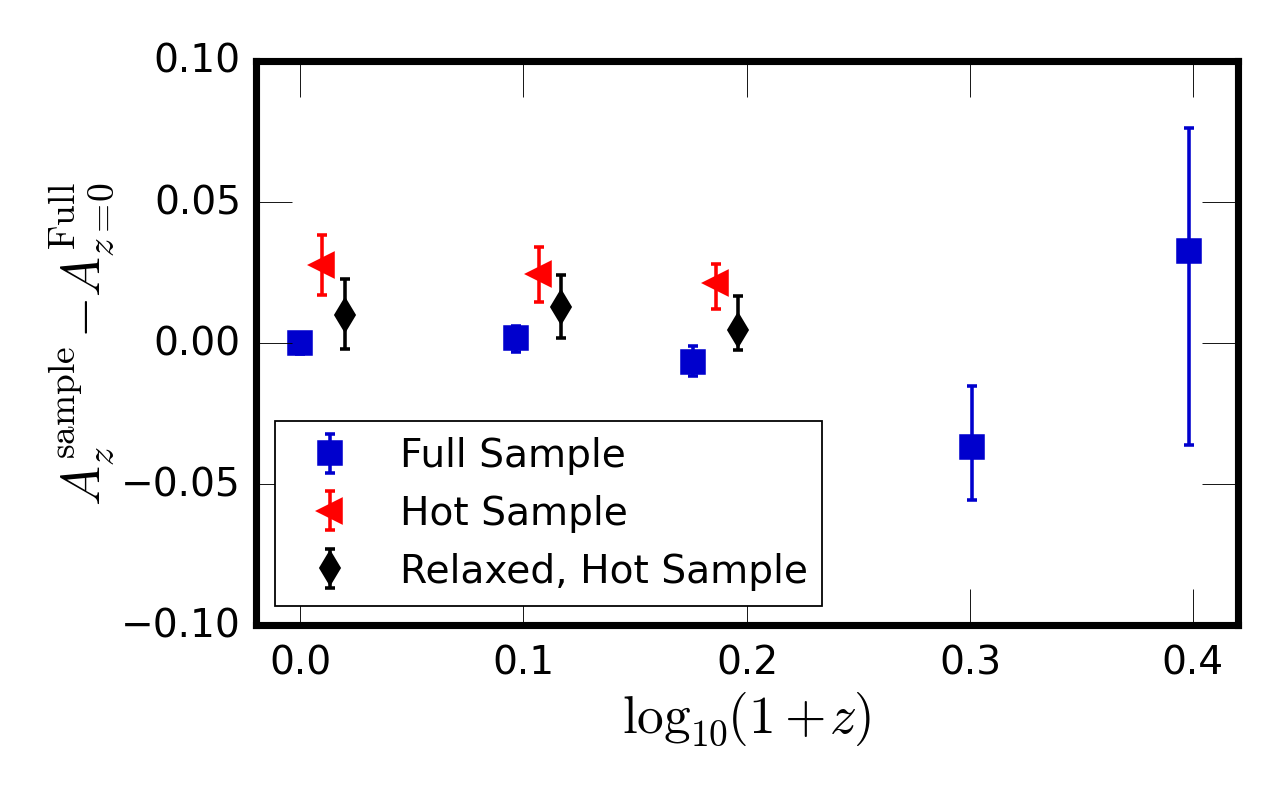}
 \caption{Evolution of the normalization of the spectroscopic temperature-total mass relation when the effective non-thermal support temperature is included. All three samples show negligible evolution with redshift relative to self-similar once non-thermal pressure support is included.}
 \label{fig:TxTkinMsr}
\end{figure}

The evolution of the core-excised spectroscopic temperature-total mass scaling relations, and their scatter, for the three samples are shown in Fig. \ref{fig:TxMsr}. The normalization of the best-fit relation of the combined sample shows a minor evolution with redshift, being $15\%$ lower at $z=1$ compared to $z=0$ (bottom left panel). In the self-similar model the temperature of the ICM is related to the depth of the gravitational potential of the cluster, under the assumption of hydrostatic equilibrium. Previous simulation work has shown that the non-thermal pressure in mass-limited samples grows with redshift due to the increasing importance of mergers and resulting incomplete thermalisation \citep{Stanek2010,LeBrun2016}. Therefore, clusters increasingly violate the assumption of hydrostatic equilibrium with redshift and require a lower temperature at a fixed mass to balance gravitational collapse, which leads to a normlization that decreases with redshift compared to self-similar. The effective temperature of the non-thermal pressure can be estimated via
\begin{equation}
 T_{\rm{kin}}=\left(\frac{\mu m_{\rm{p}}}{k_{\rm{B}}}\right)\sigma_{\rm{gas}}^2\,
\end{equation}
where $\sigma_{\rm{gas}}$ is the 1D velocity dispersion of the gas particles, $\mu=0.59$ is the mean molecular weight, $m_{\rm{p}}$ is the mass of the proton and $k_{\rm{B}}$ is the Boltzmann constant. Fig. \ref{fig:TxTkinMsr} shows the evolution of the temperature-mass normalization once this effective kinetic temperature has been added to the spectral temperature. For all three samples the addition of the kinetic temperature results in a normalization that shows significantly reduced evolution with respect to self-similar.

\begin{figure}
 \includegraphics[width=\columnwidth]{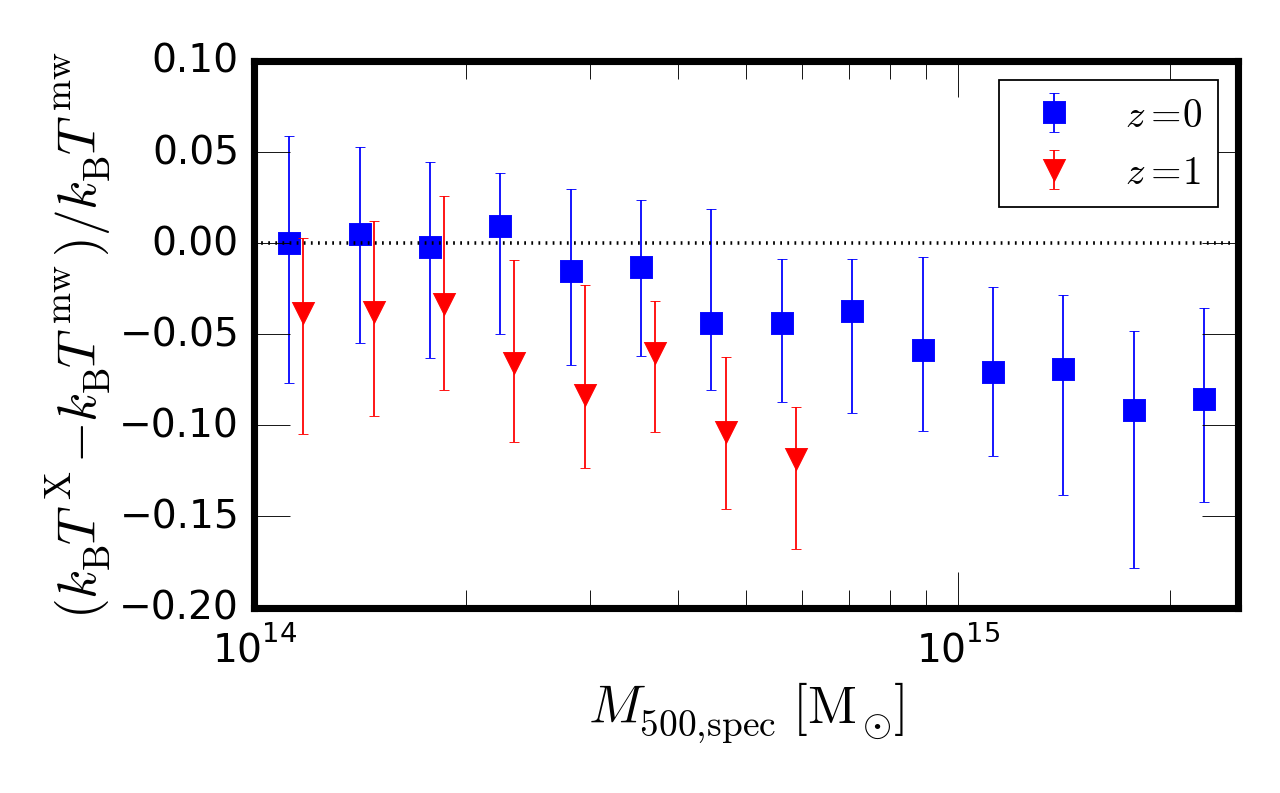}
 \caption{Plot of fractional difference between the spectroscopic and mass-weighted temperature estimates as a function of $M_{500}$ for the combined sample at $z=0$ (blue squares) and $z=1$ (red triangles). Error bars show $68\%$ of the population.}
 \label{fig:Tx_bias}
\end{figure}

The normalizations of the best-fit relations for the hot cluster and the relaxed hot samples are slightly higher than for the combined sample, but they show a similar trend with redshift that is removed when the kinetic temperature is included. The higher normalization occurs because, again, the hot sample has a flatter slope with mass. This flatter slope is driven by two processes. First, non-thermal pressure support becomes more important in higher mass clusters at a fixed redshift, as they have had less time to thermalise, and this lowers their temperatures. Second, we find that the bias between the spectroscopic and mass-weighted temperatures increases mildly with mass. This does not appear to be caused by cold clumps due to the SPH method, but is due to the presence of cooler gas in the outskirts of massive clusters that is hotter than the $0.5\,\rm{keV}$ lower limit, contributing to the X-ray spectrum, and biasing the measured temperature low for the most massive clusters. Fig. \ref{fig:Tx_bias} shows the fractional difference between the spectroscopic and mass-weighted core-excised temperatures as a function of mass. Similar to \citet{Biffi2014}, we find that for low-mass clusters the spectroscopic temperature estimate agrees well with the mass-weighted estimate at $z=0$. However, as cluster mass increases we find that the spectroscopic estimate is increasingly biased low compared to the mass-weighted estimate. This will also impact the hydrostatic mass estimate of the cluster and we refer the reader to \citet{Henson2016} for a more in-depth study. Both of these effects lead to a flattening of the slope with mass and a higher normalization for the hot samples. We note that removing the most disturbed clusters produces a marginal decrease in the normalization of the relation, which is due to the steeper slope yielding a lower normalization at the pivot point.

We find the slope of the best-fit relation for the combined sample to be $\alpha=0.58\pm0.01$ at $z=0$. This is in good agreement with the slope found by previous simulation work, where values of $0.55\pm0.01$ \citep{Short2010}, $0.576\pm0.002$ \citep{Stanek2010}, $0.54\pm0.01$ \citep{Planelles2014}, $0.56\pm0.03$ \citep{Biffi2014}, $0.60\pm0.01$ \citep{Pike2014} and $0.58$ \citep{LeBrun2016} were found. All of these are in agreement with the observed temperature-total mass relation found for volume-limited samples, with values of $0.58\pm0.03$ for a sample of clusters observed with \textit{XMM} \citep{Arnaud2007} and $0.56\pm0.07$ for a sample of low-redshift clusters \citep{Giles2015}. We note that a caveat to these comparisons is the differing mass ranges will alter the slope as the relation is not a perfect power law. All of these relations are slightly flatter than the predicted self-similar slope of $2/3$ due to non-thermal pressure support and temperature bias.

\begin{figure*}
 \includegraphics[width=\textwidth]{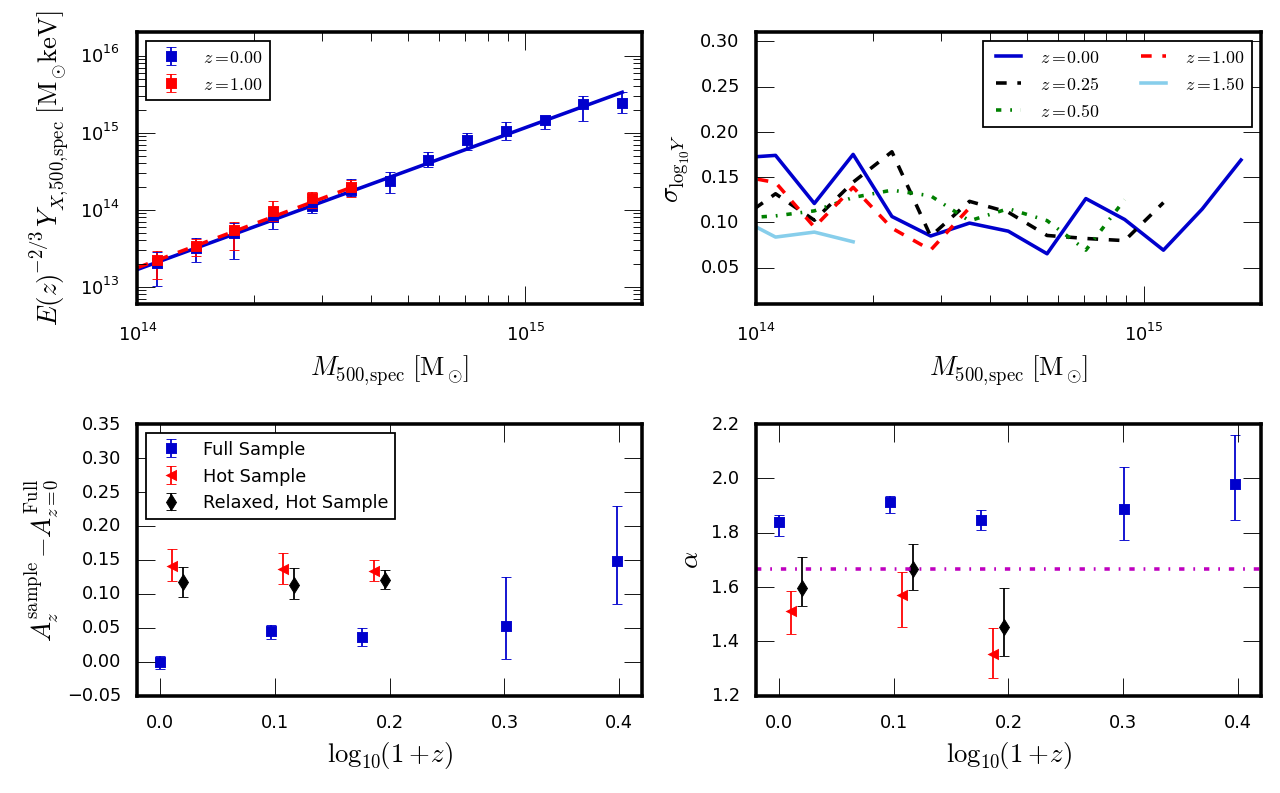}
 \caption{Evolution of the X-ray analogue $Y_{\rm{X}}$ signal-total mass scaling relation for the three samples as a function of redshift. The panels are arranged as described in Fig. \ref{fig:MgMsr}.}
 \label{fig:YxMsr}
\end{figure*}

Selecting only hot clusters produces a best-fit relation with a slope of $0.51\pm0.04$, flatter than the combined relation. The best-fit slope of $0.55^{+0.06}_{-0.03}$ for the relaxed subset, is compatible with the combined sample. The slope of the relaxed subset is compatible with the slope found by \citet{Mantz2016} of $0.66\pm0.05$ and the slope of $0.65\pm0.04$ found by \citet{Vikhlinin2009} for relaxed clusters. However, we note that our relaxation criteria only remove the most disturbed objects, as opposed to the criteria of \citet{Mantz2015} which select the most relaxed objects. Therefore, we would likely recover a steeper slope with stricter relaxation criteria. Both samples are equally affected by the spectroscopic temperature being biased low. The slopes of the hot sample and the relaxed subset show no clear trend with redshift. 

The temperature-mass scaling relations shows very low scatter, which is independent of both mass and redshift. The average scatter across all mass bins is $\sigma_{\log_{10}Y}=0.046$, $0.045$ and $0.039$ for the combined sample, hot sample and relaxed subset, respectively, at $z=0$. These values are consistent with the values found by both observations and previous simulations \citep{Arnaud2007,Giles2015,Stanek2010,Short2010}.

\subsection{$Y_{\rm{X}}$- Mass}
The power-law fits to the X-ray analogue of the integrated SZ effect-total mass relations for the three samples, and their scatter, are shown in Fig. \ref{fig:YxMsr}. The X-ray analogue signal, $Y_{\rm{X}}$, is the product of the core excised spectral temperature and the gas mass, and the relation should reflect the combination of the two previously presented relations. We indeed find this to be the case. For the combined sample, the decreasing temperature-total mass normalization with increasing redshift offsets the increasing gas mass-total mass normalization, producing almost no evolution of the normalization for the $Y_{\rm{X}}$-total mass relation. The same trend was found by \citet{LeBrun2016}. Therefore, the normalization evolves in a close to self-similar manner.

Selecting a sample of hot clusters or a relaxed subset of them leads to higher overall normalization of the best-fit relation. This is mainly due to the reduced impact of AGN feedback on the gas mass-total mass relation, which flattens the relation and leads to a higher normalization at the pivot. Both samples agree very well with the predicted self-similar evolution of the normalization of the relation, with the normalization of the relaxed subset changing by less than one percent between $z=0$ and $z=0.5$.

The slope of the $Y_{\rm{X}}$-total mass relation is simply the sum of the slopes of the temperature-mass and gas mass-total mass relations and for the combined sample the slope is significantly steeper than the $5/3$ value predicted by self-similar theory. We find a value of $\alpha=1.84^{+0.02}_{-0.05}$ at $z=0$. The slope of our best-fit relation is consistent with those of previous simulations, who found values of $1.78\pm0.01$ \citep{Short2010}, $1.73\pm0.01$ \citep{Planelles2014} and $1.89$ \citep{LeBrun2016}. Our result is also in agreement with the observational value found by \citet{Arnaud2007} of $1.82\pm0.1$ using the REXCESS cluster sample. The physical reason for the steeper slope is that gas is preferentially removed from lower mass clusters by feedback. In response to gas expulsion the remaining gas increases in temperature, offsetting some of the losses, but the loss of gas dominates and steepens the relation. The value of the slope for the best-fit relation is approximately constant with redshift, within the uncertainty of the fits.

\begin{figure*}
 \includegraphics[width=\textwidth]{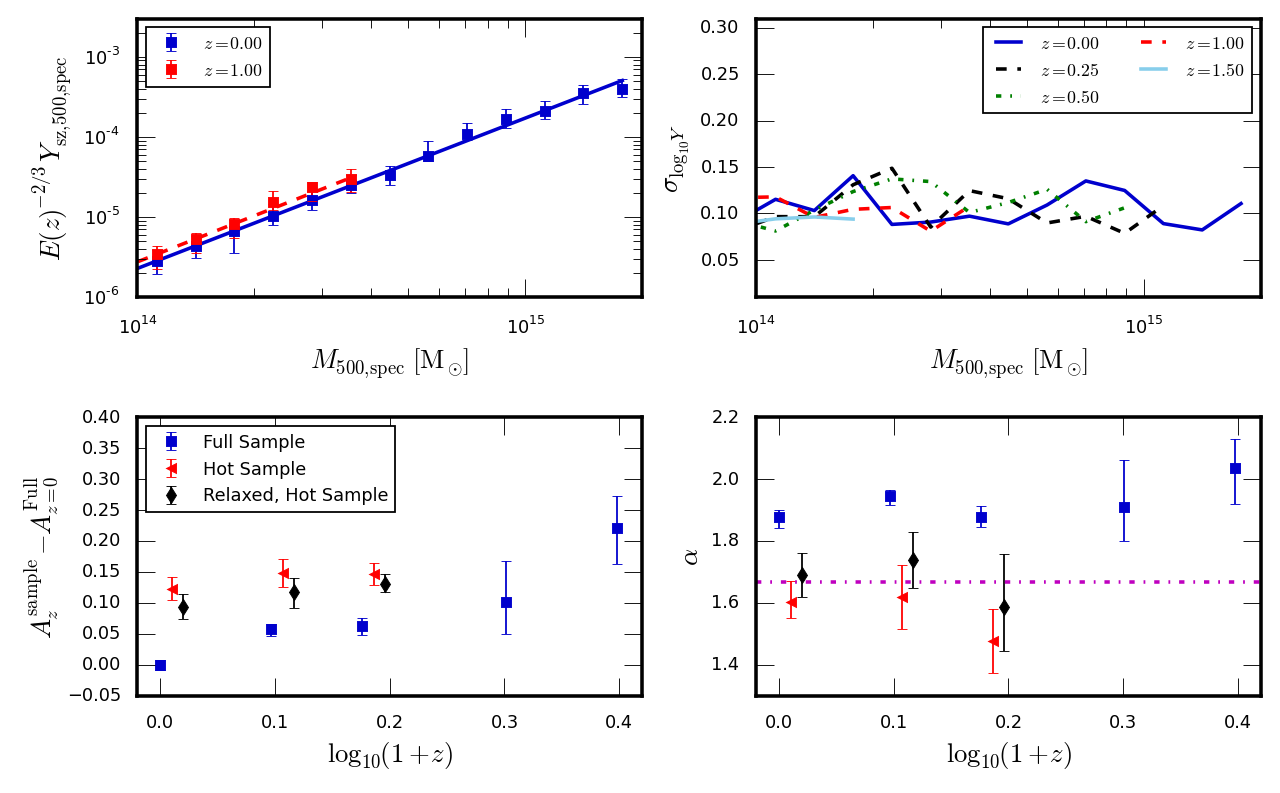}
 \caption{Evolution of the integrated Sunyaev-Zel'dovich signal-total mass scaling relation for the three samples as a function of redshift. The panels are arranged as described in Fig. \ref{fig:MgMsr}.}
  \label{fig:YszMsr}
\end{figure*}

Selecting a sample of hot clusters leads to a significant flattening of the slope of the relation, slightly flatter than the self-similar prediction of $5/3$. With the gas mass-total mass relations of the hot sample and relaxed subset being very close to self-similar, the shallower than self-similar slope is due to the temperature-mass relation. The best-fit slope of both samples shows no significant trend with redshift.

The scatter about the best-fit relation is independent of both mass and redshift for all three samples, but it is noisy. We find an average value of $0.12$ at $z=0$ for the scatter for the combined sample, $0.11$ for the hot cluster sample and $0.08$ for the relaxed subset. These values are larger than those found previously for both simulations, where values of $0.04$ \citep{Short2010}, $0.08$ \citep{Planelles2014} and $0.04$ \citep{LeBrun2016} were found, and observations, where a value of $0.04$ was found for a sample of clusters observed with \textit{XMM} \citep{Arnaud2007}.

\subsection{$Y_{\rm{SZ}}$-Total Mass}
The integrated SZ effect-total mass relations for the three samples are shown in Fig. \ref{fig:YszMsr}. Both the integrated SZ signal and its X-ray analogue measure the total energy of the hot gas in the ICM, however the SZ signal depends on the mass-weighted temperature rather than the X-ray spectral temperature. Our best-fit relation for the combined sample shows a mild evolution with redshift, with clusters at $z=1$ yielding an integrated signal that is $27\%$ higher than clusters at $z=0$ for a fixed mass. The evolution reflects the evolution in the gas mass-total mass relation. The increased evolution of its normalization compared to its X-ray analogue suggests that the normalization of the mass-weighted temperature evolves more self-similarly then the spectroscopic X-ray temperature and is indeed confirmed by the study of the mass-weighted temperature-total mass relation.

Selecting a sample of hot clusters or a relaxed subset of them significantly reduces the evolution in the normalization. The normalization of both samples, within the uncertainty of the fits, evolves in agreement with the self-similar prediction. Selecting a hot sample leads to a $25\%$ higher normalization than the combined sample at $z=0$, due to the flatter slope of the gas mass-total mass relation yielding a flatter $Y_{\rm{SZ}}$ slope and a higher normalization at the pivot point.

The best-fit relation for the combined sample produces a slope of $\alpha=1.88^{+0.02}_{-0.04}$ at $z=0$, which is significantly steeper than the $5/3$ value predicted by the self-similar model. The value for the slope of the relation is consistent with previous values from both simulations, where values of $1.825\pm0.003$ \citep{Stanek2010}, $1.71\pm0.03$ \citep{Battaglia2012}, $1.74\pm0.01$ \citep{Planelles2014}, $1.70\pm0.02$ \citep{Pike2014}, $1.68\pm=0.05$ \citep{Yu2015} and $1.94$ \citep{LeBrun2016} have been found, and observations, where $1.79\pm0.08$ was found for the Planck clusters \citep{Planck2014XX} and $\alpha=1.77\pm0.35$ was found for the clusters in the $2500\,\rm{deg}^2$ SPT survey. The steeper than self-similar slope is the result of the gas mass-total mass relation having a steeper slope. We find that the slope of the relation is independent of redshift.

\begin{figure*}
 \includegraphics[width=\textwidth]{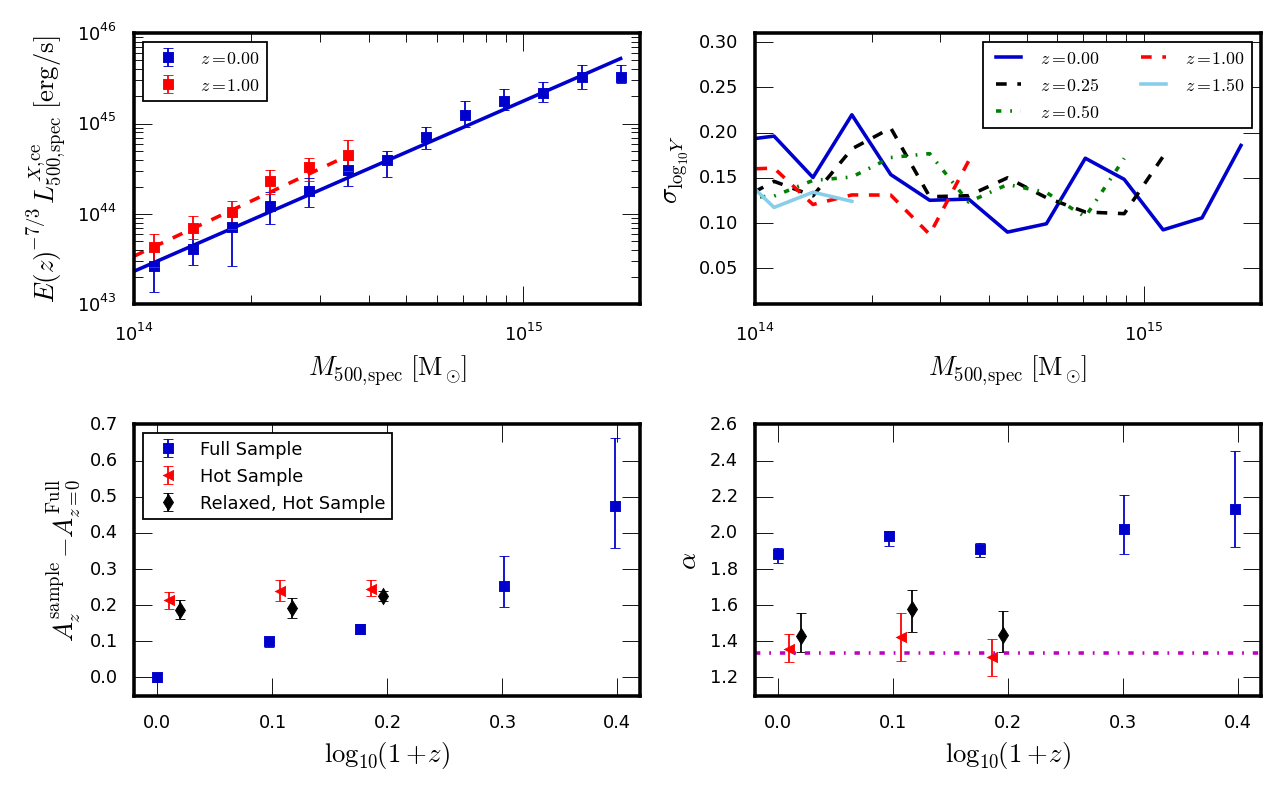}
 \caption{Evolution of the core-excised bolometric X-ray luminosity-total mass scaling relation as a function of redshift for the three samples of clusters. The panels are arranged as described in Fig. \ref{fig:MgMsr}.}
 \label{fig:LxMsr}
\end{figure*}

The best-fit slopes of the hot cluster sample and the relaxed subset are consistent with the slope predicted by self-similar theory. The slopes of both samples are consistent with no evolution.

The scatter of the clusters about the best-fit relation shows no trend with either mass or redshift for all three samples. We find an average scatter of $\sigma_{\log_{10}Y}=0.10$, $0.10$ and $0.09$ for the combined, hot and relaxed samples, respectively, at $z=0$. This is larger than the scatter reported by previous simulations, where \citet{Battaglia2012}, \citet{Pike2014}, \citet{Planelles2014} and \citet{LeBrun2016} found values of $0.06$, $0.03$, $0.07$ and $0.04$ respectively, but in reasonable agreement with the values of $0.12\pm0.03$ and $0.08$ observed by \citet{Yu2015} and \citet{Planck2014XX} respectively.

\subsection{Bolometric X-ray Luminosity-Total Mass}
Fig. \ref{fig:LxMsr} shows the core-excised bolometric X-ray luminosity-total mass scaling relations for the three samples and their evolution with redshift. The normalization of the best-fit relation for the combined sample shows significant evolution with redshift, being $80\%$ higher at $z=1$ compared to $z=0$. The same physics driving the gas mass-total mass relation, increased binding energy, is driving the departure from self-similar. The X-ray emission of a cluster is particularly sensitive to the thermal structure of the ICM, which depends on processes such as radiative cooling and feedback. Therefore, it is not surprising that the luminosity-mass relation shows significantly more evolution than other observable-mass relations.

Selecting a sample of hot clusters significantly reduces the evolution in the normalization. Both the hot sample and the relaxed subset have a normalization that is $\approx60\%$ larger at $z=0$ compared to the combined sample. The deeper potentials of more massive clusters reduces the impact of the AGN feedback and flattens the relation. This flattening leads to a higher luminosity at the pivot point. The normalizations of the best-fit relations for both the hot sample and its relaxed subset show very minor evolution, which is consistent with the self-similar prediction.

\begin{figure*}
 \includegraphics[width=\textwidth]{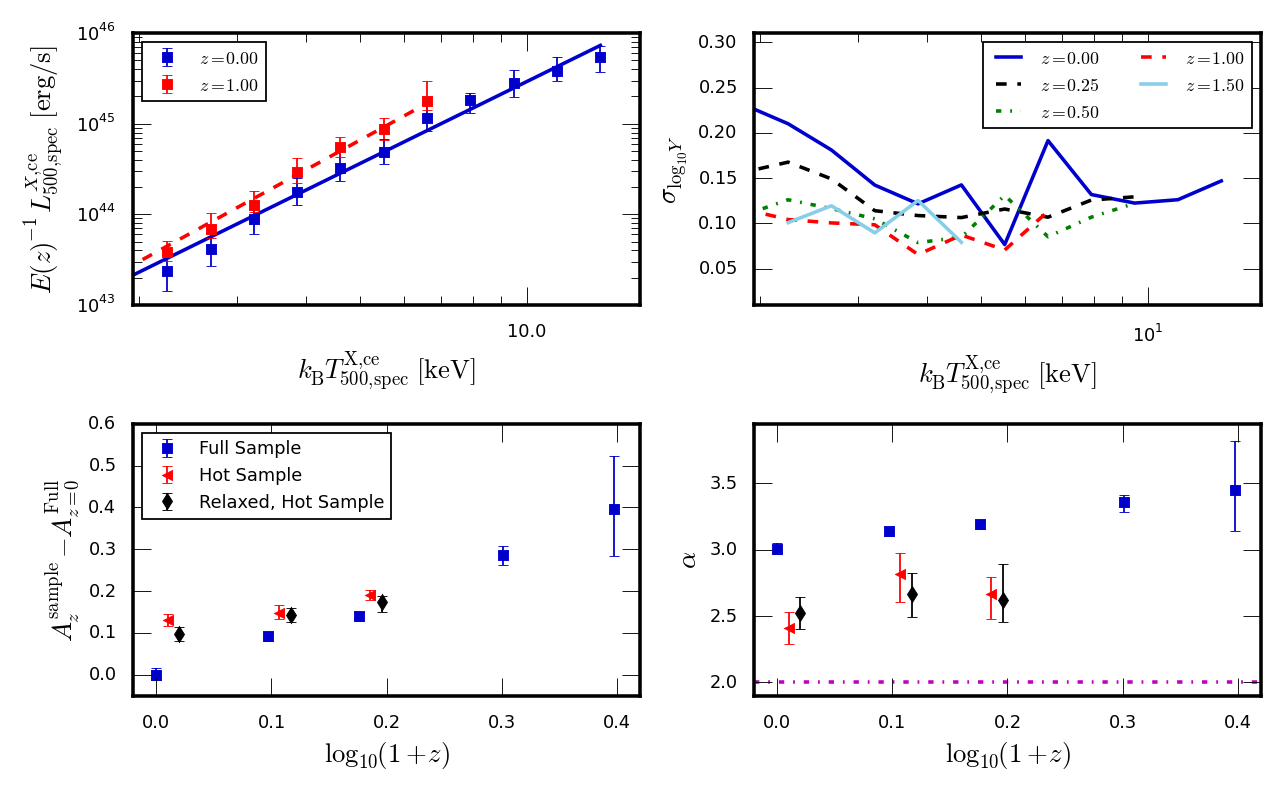}
 \caption{Evolution of the bolometric X-ray luminosity-X-ray temperature scaling relation for the three samples as a function of redshift. The panels are arranged as described in Fig. \ref{fig:MgMsr}.}
 \label{fig:LxTxsr}
\end{figure*}

The slope of the best-fit relation for the combined sample is significantly steeper than the $4/3$ slope predicted by self-similar theory. At $z=0$ we find a slope of $\alpha=1.88^{+0.03}_{-0.05}$ for the combined sample. This steepening is driven by AGN feedback being more effective in lower mass clusters. The slope at $z=0$ is in reasonable agreement with the slopes found in volume-limited observational samples, such as \citet{Pratt2009} who found a slope of $1.80\pm0.05$ for the REXCESS sample and \citet{Giles2015} who found a slope of $2.14\pm0.21$ for a sample of $34$ low-redshift clusters. Previous simulation work by \citet{Short2010}, using the semi-analytic feedback model of the Millennium Gas project, found a bolometric luminosity-total mass slope of $1.77\pm0.03$ and \citet{Stanek2010}, using the preheating model of the Millennium Gas project, found a slope of $1.87\pm0.01$. \citet{Biffi2014} found a slope of $1.45\pm0.05$ for the MUSIC simulations. The slope of the best-fit relation for the combined sample is approximately independent of redshift, with a very mild steepening of the slope with redshift occurring due to the reduction in fitting range with increasing redshift.

The slopes of the best-fit relation follow the same trend as the gas mass-total mass relation, with the hot sample and its relaxed subset producing shallower slopes that are in much better agreement with self-similar theory. Our best-fit slope is consistent with the observational result of \citet{Mantz2016}, who found a self-similar slope for the core-excised luminosity-total mass relation for a sample of $40$ relaxed clusters with $k_{\rm{B}}T\geq\,5\,\rm{keV}$.

The scatter about the best fit relation is approximately independent of both mass and redshift for all three samples, although it is relatively noisy. Averaging the scatter for the combined sample across all mass bins produces a value of $\sigma_{\log_{10}Y}=0.15$. This is in reasonable agreement with the scatter found in low-redshift observational samples, where \citet{Pratt2009} find a value of $0.17\pm0.03$ and \citet{Giles2015} find a value of $0.22\pm0.03$. Selecting hot clusters and a relaxed subset produces a small reduction in scatter about the best-fit relation with values of $0.12$ and $0.11$ respectively.

\subsection{X-ray Luminosity-Temperature}
Finally, we study the redshift evolution of the X-ray luminosity-spectroscopic temperature relation. Both quantities of the luminosity-temperature scaling relation are observable, with the temperature tracing the depth of the potential of the cluster. This makes it a useful relation to study the impact of non-gravitational physics. In Fig. \ref{fig:LxTxsr} we plot the bolometric X-ray luminosity-spectroscopic temperature scaling relation for the three samples of clusters. The normalization of the best-fit relation for the combined sample shows significant evolution with redshift relative to self-similar. Clusters with a temperature of $6\,\rm{keV}$ at $z=1$ have a luminosity $94\%$ greater than clusters with the same temperature at $z=0$. This evolution can be thought of as being due to a combination of the evolution of the temperature-mass and luminosity-mass relations. The decreasing temperature-mass normalization and increasing luminosity-mass normalization with redshift combine to yield a significant evolution of the luminosity-temperature normalization relative to self-similar.

Selecting a sample of hot clusters, or a relaxed subset of them, reduces the evolution, but there is still a mild evolution in the normalization. Hot clusters at a fixed temperature at $z=0.5$ are $\approx15\%$ more luminous than those at $z=0$. Combining equations (\ref{eq:T-M}) and (\ref{eq:L-M}), but allowing the slope of the relations to vary from their self-similar values yields
\begin{equation}
 \label{eq:Lx-Txevo}
 L_{\rm{X},\Delta}^{\rm{bol}}\propto T^{\alpha_{\rm{LM}}/\alpha_{\rm{TM}}}E^{7/3-2\alpha_{\rm{LM}}/3\alpha_{\rm{TM}}}(z)\,,
\end{equation}
where $\alpha_{\rm{LM}}$ and $\alpha_{\rm{TM}}$ are the slopes of the luminosity-mass and temperature-mass relations respectively. Hence, deviations of their slopes from self-similar leads to evolution of the normalization of the luminosity-temperature relation that is not self-similar. With the luminosity-mass relation being self-similar for the hot cluster sample and its relaxed subset, the evolution of the normalization is being driven by the flatter than self-similar slope of the temperature-mass relation, which is due to the increased importance of non-thermal pressure support and the increasingly biased spectroscopic temperatures of more massive clusters.

We find a slope of $\alpha=3.01\pm0.04$ for the best-fit relation at $z=0$. This is significantly steeper than the slope of $2$ predicted by self-similar theory. However, this value is reasonable agreement with previous simulation work, $3.30\pm0.07$ \citep{Short2010}, and those found by observations, $2.95\pm0.15$ for the REXCESS sample \citep{Pratt2009} and $\alpha=3.63\pm0.27$ for a sample of $114$ clusters observed with \textit{Chandra} \citep{Maughan2012}. It is clear from equation (\ref{eq:Lx-Txevo}) that the slope of the relation depends on the slopes of the luminosity-mass and temperature-mass relations. The steeper than expected slope for the combined sample is due to the combined effects of AGN feedback on the luminosity slope and non-thermal pressure support and temperature bias on the temperature slope, both of which lead to a steepening of the relation compared to the self-similar prediction. We find that the best-fit relation steepens slightly with redshift, increasing to $3.35\pm0.07$ at $z=1$. This evolution is due to the removal of high-mass objects with redshift.

The best-fit slope of the hot cluster sample and the relaxed subset are flatter than the combined relation with slopes of $2.41\pm0.12$ and $2.53\pm0.13$. This is still significantly steeper than the slope predicted by self-similar theory, but in good agreement with the slope of $2.44\pm0.43$ observed by \citet{Maughan2012} for their relaxed cool core cluster sample. With both samples exhibiting self-similar slopes for the luminosity-mass relations, the deviation from self-similarity is being driven by their temperature-mass relations.

The scatter about the best-fit relation demonstrates a trend with both temperature and redshift. Although somewhat noisy, the scatter appears to increase with decreasing temperature. The average scatter at $z=0$ for the combined sample is $\sigma_{\log_{10}Y}=0.14$ . This scatter is consistent with the simulations of \citet{Short2010}, who found a scatter $0.10$, and the intrinsic observational scatter of $0.12$ found by \citet{Pratt2009}. However, it is significantly lower than the scatter of $0.29$ found by \citet{Maughan2012}. The scatter reduces for the hot cluster sample and the relaxed subset to $0.11$ and $0.10$ respectively.

\subsection{Summary}
Overall, the scaling relations of the combined sample show good agreement with previous work, both simulations and observations. Departures from self-similarity are driven by the increased efficiency of gas expulsion by AGN feedback in clusters with shallower potentials, due to being less massive or forming at a lower redshift; the increased contribution of non-thermal pressure that supports the ICM against gravity in more massive clusters or those at higher redshifts; and the increase in the spectroscopic temperature bias for the most massive clusters. The MACSIS sample enabled the scaling relations to be studied to higher redshifts, as their progenitors are still clusters at high redshift, and the examination of the impact of selecting a sample of hot clusters on the evolution of the scaling relations. This demonstrated that massive clusters are more self-similar and evolve more self-similarly with redshift compared to the overall cluster population, as the efficiency of gas expulsion by AGN feedback is reduced due to their deeper potentials. However, it also highlighted that non-thermal pressure support becomes more important in these clusters and that their spectroscopic temperatures are biased low.

\section{Evolution of gas profiles}
\label{sec:gasprofs}
Most of the scaling relations of hot, and therefore massive, clusters evolve in a way that is consistent with the predictions of the self-similar model. However, the combined sample showed significant deviations from the self-similar model due to the impact of non-gravitational processes. To further understand the differences between the samples in the evolution of their scaling relations, we now examine the gas profiles of the different cluster samples. To enable a quantitative comparison with the observational data requires us to compare like-with-like. Therefore, we restrict the mass range of the combined sample to $2.0\times10^{14}\,\mathrm{M}_{\odot}\leq M_{500,\mathrm{spec}}\leq1.0\times10^{15}\,\mathrm{M}_{\odot}$, yielding a sample with a median mass of $2.44\times10^{14}\,\mathrm{M}_{\odot}$. We compare this to the REXCESS cluster sample which has a median mass of $2.68\times10^{14}\,\mathrm{M}_{\odot}$ and a sample of clusters from \citet{Giles2015} with a median mass of $5.43\times10^{14}\,\mathrm{M}_{\odot}$. Although this mass matching does not account for selection effects, it should allow for a quantitative comparison. We do not alter the hot sample or the relaxed subset. We factor out the expected self-similar evolution in the profiles by dividing by the appropriate quantity, e.g. $\rho_{\rm{crit}}$, $k_{\rm{B}}T_{500}$, $P_{500}$ or $K_{500}$. We define these quantities as
\begin{equation}
 \rho_{\rm{crit}}(z)\equiv E^2(z)\frac{3H^2_0}{8\pi G}\,,
\end{equation}
\begin{equation}
 k_{\rm{B}}T_{500}=\frac{GM_{500}\mu m_{\rm{p}}}{2r_{500}}\,,
\end{equation}
\begin{equation}
 P_{500}=500f_{\rm{b}}k_{\rm{B}}T_{500}\frac{\rho_{\rm{crit}}}{\mu m_{\rm{p}}}\,,
\end{equation}
\begin{equation}
 K_{500}=\frac{k_{\rm{B}}T_{500}}{\left(500f_{\rm{b}}(\rho_{\rm{crit}}/\mu_{\rm{e}}m_{\rm{p}})\right)^{2/3}}\,,
\end{equation}
where $H_0$ is the Hubble constant, $G$ is the gravitational constant, $\mu_{\rm{e}}$ is the mean atomic weight per free electron and $f_{\rm{b}}=\Omega_{\rm{b}}/\Omega_{\rm{m}}$ is the universal baryon fraction. Therefore, any changes in the profiles are due to non-gravitational physics, such as AGN feedback or non-thermal pressure support.

\subsection{Density profiles}
\begin{figure}
 \includegraphics[width=\columnwidth]{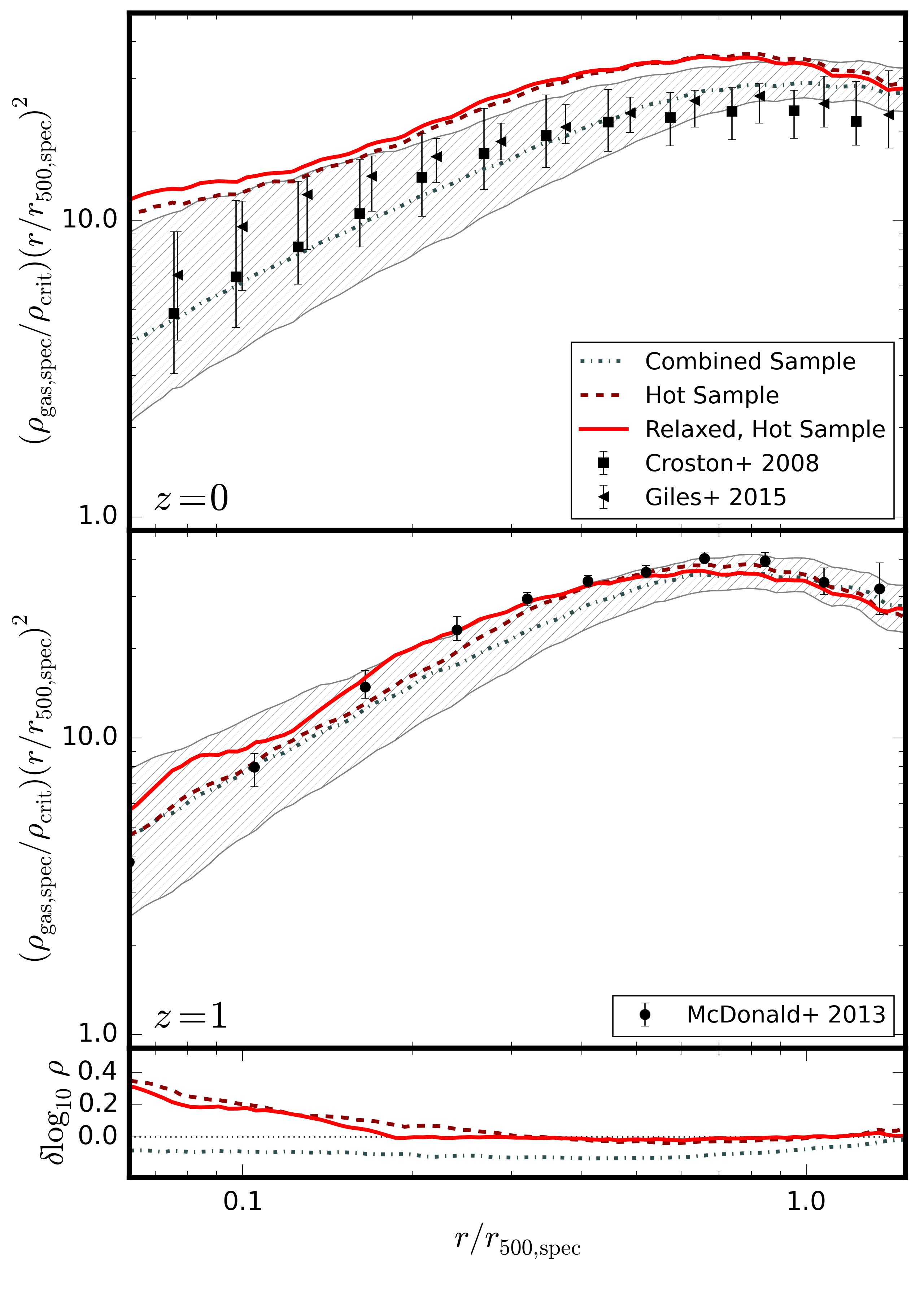}
 \caption{Median gas density profiles for the combined (grey dash-dot), hot (dark red dashed) and relaxed hot (red solid) samples at $z=0$ (top panel) and $z=1$ (middle panel), scaled by $(r/r_{500,\rm{spec}})^2$ to reduce dynamic range. The grey hatched region shows the $16^{\rm{th}}$ to $84^{\rm{th}}$ percentiles of the combined sample profile. Overlaid as black squares, triangles and circles are the median observed profiles from the REXCESS sample \citep{Croston2008}, a sample of low redshift clustes observed with \textit{Chandra} \citep{Giles2015} and a high-redshift, SPT-selected sample \citep{McDonald2013,McDonald2014} respectively, with the error bars showing the $16^{\rm{th}}$ and $84^{\rm{th}}$ percentiles. The bottom panel shows the $\log_{10}$ of the ratio of the profiles at $z=0$ and $z=1$ for each sample.}
 \label{fig:gas_prof}
\end{figure}

The three-dimensional dimensionless density profiles for the three cluster samples at $z=0$ and $z=1$ are shown in Fig. \ref{fig:gas_prof}. We have scaled the profiles by $r^2$ to reduce the dynamic range. At $z=0$, we compare the median profile of the combined sample with the observed median profiles from \citet{Croston2008} for the REXCESS sample and \citet{Giles2015} for a sample of low-redshift clusters observed with \textit{Chandra}. The combined sample shows good agreement with the observed profiles and has similar intrinsic scatter. Beyond a radius of $0.15r_{500,\rm{spec}}$ the median profiles of the hot sample and its relaxed subset have a similar shape as the combined sample, but the densities are higher as they are on average more massive clusters. Inside this radius the profiles of both samples have a shallower gradient compared to the combined sample. This is caused by the accretion of low-entropy, high-density gas that sinks to the centre of the cluster potential, becoming increasingly important below $z=1$ \citep{Power2014}. This effect is not offset in massive clusters by the AGN feedback and so their density profiles have a shallower gradient in the core. We note that this effect can potentially impact the relations we presented in section \ref{sec:screlations}. However, we presented core-excised temperatures and luminosities, which should minimise any bias introduced by the accretion of poorly mixed gas.

At $z=1$, we compare the median density profiles of the three samples to the observed profile from \citet{McDonald2013}, which has been derived from a sample of $40$ clusters with a mean redshift of $z=0.82$. These clusters were selected from the SPT $2500\,\rm{deg}^2$ survey catalogue and observed with \textit{Chandra}. There is a reasonable agreement between the combined sample's median profile and the observations, but the observations are higher between $0.2-1.0r_{500}$. The observed profile is in better agreement with the median profiles of the hot sample and its relaxed subset. This suggests that the observed clusters are more representative of more massive objects at $z=1$. There is better agreement between the density profiles of the three samples at $z=1$ because the mass cut of $M=10^{14}\,\rm{M}_{\odot}$ causes the samples to converge with increasing redshift. Selecting relaxed hot clusters leads to a median profile that is slightly more centrally concentrated than for all hot clusters.

In the bottom panel of the plot we show the $\log_{10}$ of the ratio of the median density profile at $z=0$ and the median profile at $z=1$ for each sample. For the hot cluster sample and the relaxed subset the profiles have evolved in a self-similar way beyond $0.2\,r_{500}$, showing very little change. Inside of this radius the impact of accreting low-entropy, high-density gas that sinks to the centre of the cluster is apparent as an increase in the density profiles from $z=1$ to $z=0$. For the combined sample the difference between the two profiles shows the increase of the depth of the potential with redshift. This leads to higher densities at $z=1$ and a negative change density profile at all radii with decreasing redshift.

\subsection{Temperature profiles}
\begin{figure}
 \includegraphics[width=\columnwidth]{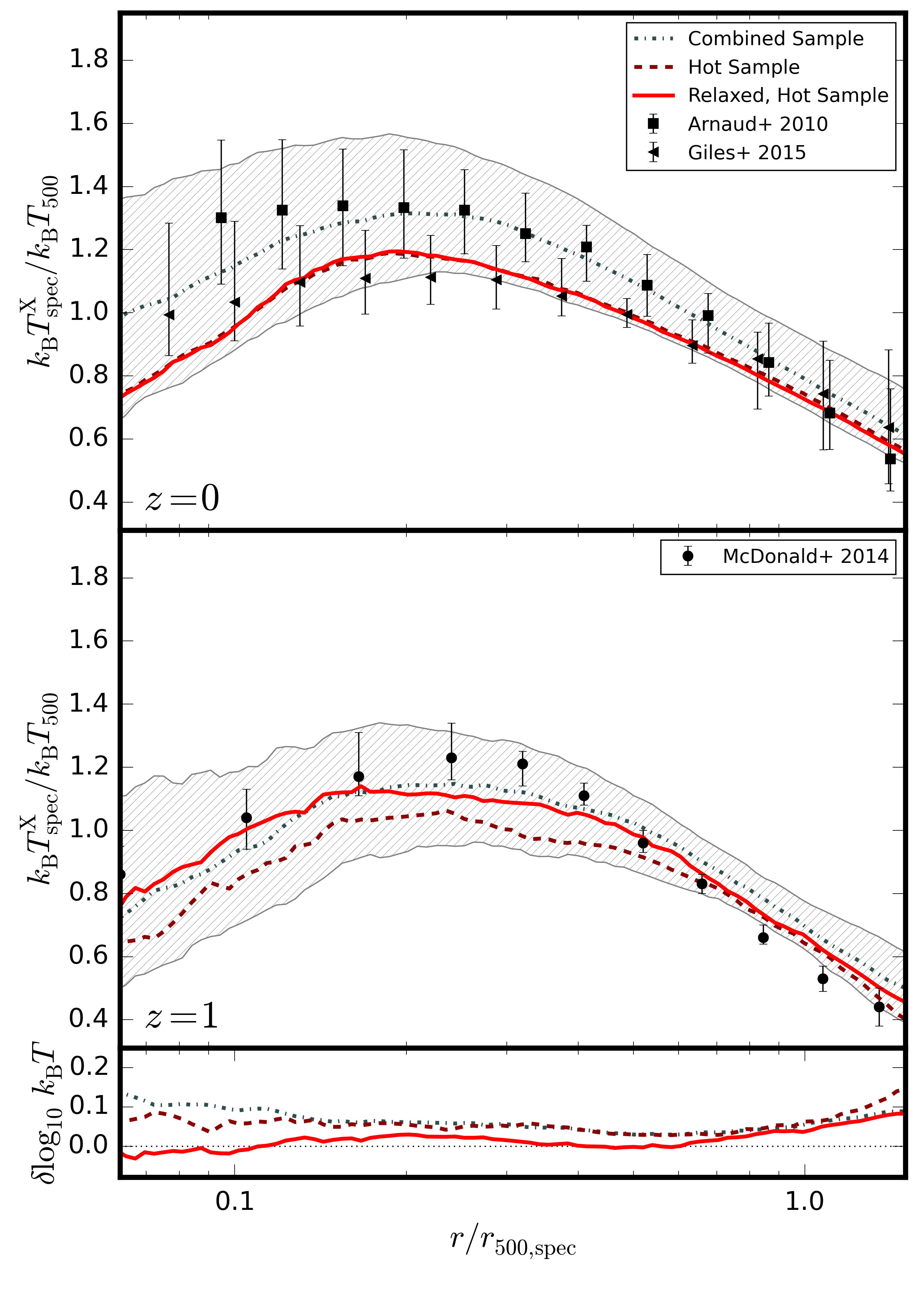}
 \caption{Median temperature profiles for the three samples. The details are the same as for Fig. \ref{fig:gas_prof}, except that the REXCESS data was taken from \citet{Arnaud2010}.}
 \label{fig:temp_prof}
\end{figure}

Fig. \ref{fig:temp_prof} shows the three-dimensional temperature profiles divided by the predicted self-similar temperature. At $z=0$ the profiles all have a similar shape, but the normalization of the combined sample is somewhat higher than those of the hot sample and its relaxed subset. This is due to the lower gas density of the combined sample, which requires a higher temperature to balance gravitational collapse. Also, there is likely to be a small effect due to the mass dependence of non-thermal pressure support, with more massive clusters having more non-thermal support and lower temperatures. The accretion of low-entropy, cold gas that sinks to the cluster core produces a steeper temperature gradient in the central profiles of the hot sample and its relaxed subset. Overlaid are the observed median temperature profiles from two cluster samples, the REXCESS sample \citep{Arnaud2010} and a sample of clusters observed with \textit{Chandra} \citep{Giles2015}. The median profile of the combined sample and its intrinsic scatter show good agreement with the observed temperature profiles and their scatter.

At $z=1$ all samples have a similar profile shape, but the hot sample has a lower normalization compared to the combined and relaxed hot sample. This is because non-thermal pressure support becomes increasingly important in clusters of a fixed mass with redshift, leading to a lower temperature in hot clusters. The relaxed sample removes the most disturbed objects with greatest level of non-thermal support, producing a higher median temperature profile. We compare to the observed median profile of \citet{McDonald2014}. The median profiles of the combined sample and the relaxed hot sample slightly under predict the observations at $0.3\,\rm{r}_{500}$ and over predict the observations at large radii, but the observed profile is within the scatter of the combined sample.

Within $r_{500,\rm{spec}}$ the median temperature profiles show significantly less evolution between the two redshifts than the density profiles. The combined and hot samples deviate from self-similarity and show an increase in temperature from $z=1$ to $z=0$ at all radii, consistent with the decreasing temperature-mass normalization with increasing redshift found in Section \ref{ssec:Tx-M}. This is because non-thermal pressure support decreases with increasing redshift. Therefore, as clusters thermalise their temperatures must increase to balance gravitational collapse, resulting in a hotter temperature profile at $z=0$ compared to $z=1$. Selecting a relaxed subset reduces the non-thermal pressure support and the median profile changes significantly less from $z=1$ to $z=0$ inside $r_{500}$.

\subsection{Pressure profiles}
\begin{figure}
 \includegraphics[width=\columnwidth]{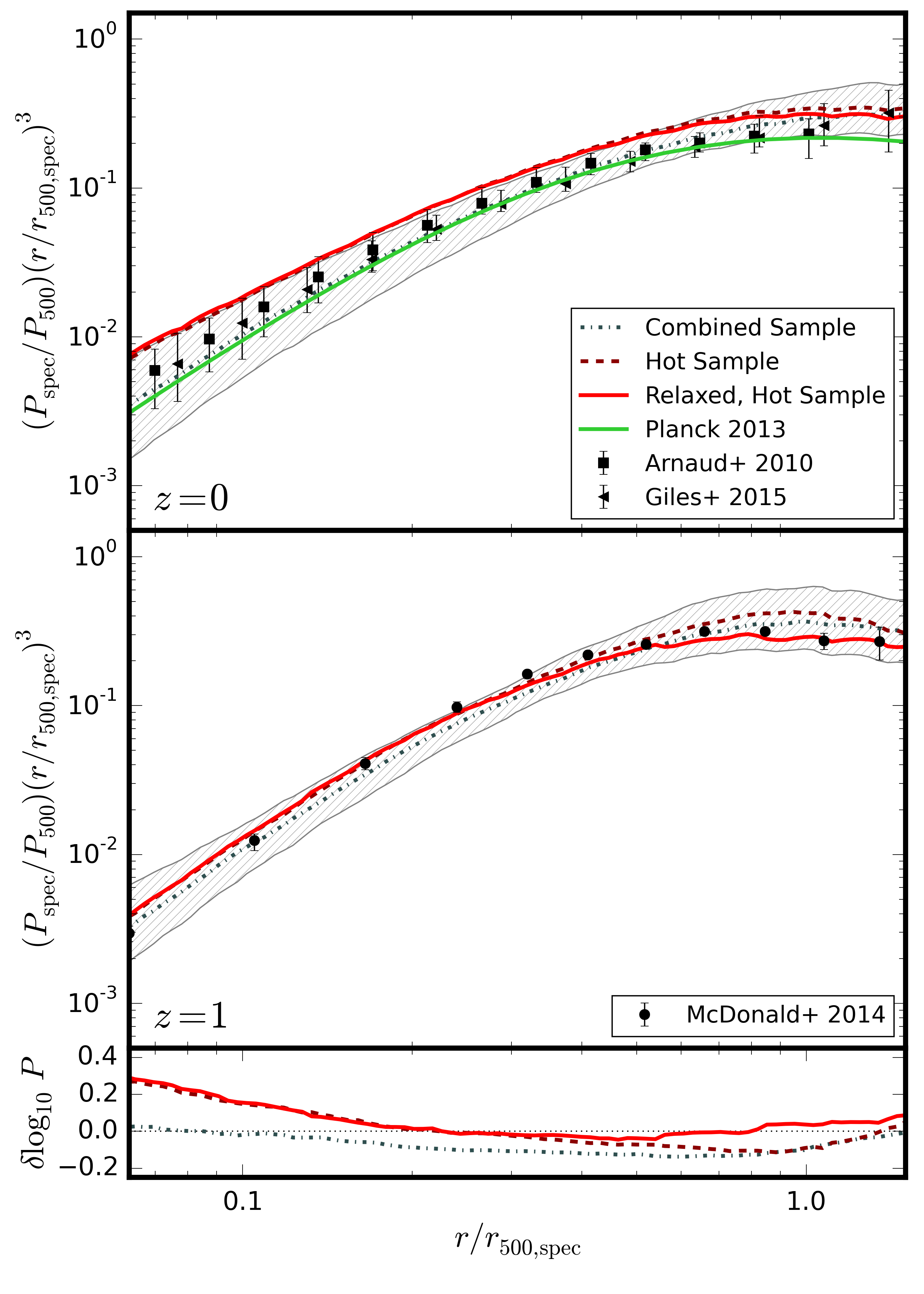}
 \caption{Median pressure profiles for the three samples. The details are the same as for Fig. \ref{fig:gas_prof}, except that the REXCESS data was taken from \citet{Arnaud2010}. The green curve shows the best-fit pressure profile from \citet{Planck2013}.}
 \label{fig:pres_prof}
\end{figure}

The dimensionless pressure profiles, scaled by $r^{3}$, of the three cluster samples are shown in Fig. \ref{fig:pres_prof}. The increased mass of the hot sample and its relaxed subset lead to median pressure profiles that are higher in the centre at $z=0$ due to their higher densities. We compare the median profiles to the observed median pressure profiles from \citet{Arnaud2010} and \citet{Giles2015} and the best-fit profile from \citet{Planck2013}. We note that the \textit{Planck} result is based on the stacked profile of nearby systems. For \citet{Giles2015} we have combined their published density and temperature profiles to produce a pressure profile for each cluster. There is good agreement between the combined sample and the observed profiles, with a slight over prediction at large radii. For comparison to the Planck best-fit parameters we fit the mean profiles of our clusters at both redshifts with a generalised Navarro-Frenk-White pressure profile \citep{NFW1997,NagaiVikhlininKravtsov2007a} of the form
\begin{equation}
 P(x)=\frac{P_{0}}{(c_{500}x)^{\gamma}\left[1+(c_{500}x)^{\alpha}\right]^{(\beta-\gamma)/\alpha}}\:.
 \label{gNFWeq}
\end{equation}
We fit a four parameter model with $\gamma=0.31$ fixed. The results are shown in table \ref{tab:Ppro_fit}.

\begin{table}
 \caption{Table showing the best-fit generalised Navarro-Frenk-White pressure profile parameters (see eq. \ref{gNFWeq}) for the combined, hot and relaxed hot samples of clusters present in this work. We fix $\gamma=0.31$.}
 \centering
 \begin{tabular}{c l r c c c}
  \hline
  $z$ & Sample & \multicolumn{1}{c}{$P_{0}$} & $c_{500}$ & $\alpha$ & $\beta$ \\
  \hline
   & \textit{Planck} & $6.41$ & $1.81$ & $1.33$ & $4.13$ \\
  \hline
  $0$ & Combined & $8.80$ & $1.56$ & $1.09$ & $4.01$ \\
   & Hot & $20.66$ & $0.52$ & $0.70$ & $6.69$ \\
   & Relaxed Hot & $24.01$ & $0.54$ & $0.69$ & $6.79$ \\
  \hline
  $1$ & Combined & $6.96$ & $0.99$ & $1.26$ & $5.84$ \\
   & Hot & $6.44$ & $0.51$ & $1.14$ & $9.44$ \\
   & Relaxed Hot & $9.28$ & $1.97$ & $1.61$ & $4.11$ \\
  \hline
 \end{tabular}
 \label{tab:Ppro_fit}
\end{table}

At $z=1$ the median profiles of the three samples are in closer agreement with each other, because the minimum mass limit of $M=10^{14}\,\rm{M}_{\odot}$ causes the samples to converge at high redshift. We compare our median pressure profiles with the observed profile of \citet{McDonald2014}. They find a median pressure profile that is in good agreement with the median profiles, but it is most consistent with the relaxed hot sample of massive clusters.

The pressure profile of the relaxed subset shows very little evolution between $z=1$ and $z=0$, except for the core where the increasing density leads to an increased pressure with decreasing redshift. The hot sample shows an increased pressure in the core with decreasing redshift, due to the increased density, but a negative change in pressure from $z=1$ to $z=0$ at larger radii. The combined sample shows a negative pressure change between $z=1$ and $z=0$ at all radii. The decreased pressure with decreasing redshift is caused by the decrease in density from $z=1$ to $z=0$.

\subsection{Entropy Profiles}
\begin{figure}
 \includegraphics[width=\columnwidth]{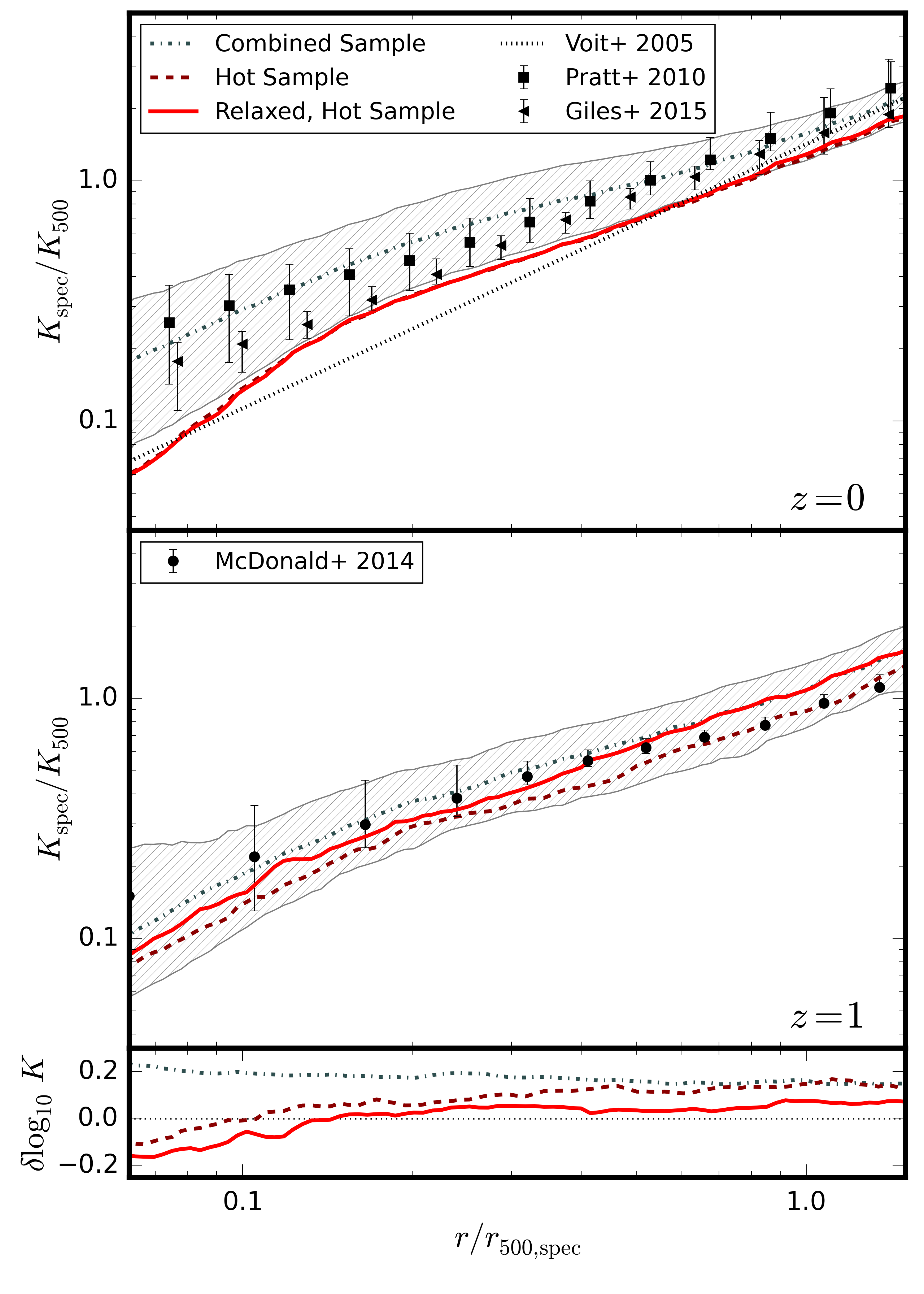}
 \caption{Median entropy profiles for the three samples. The details are the same as Fig. \ref{fig:gas_prof}, except that the REXCESS data was taken from \citet{Pratt2010}. We also show the prediction from non-radiative simulations for $z=0$ \citep{VoitKayBryan2005}.}
 \label{fig:enty_prof}
\end{figure}

The median entropy profiles are shown in the bottom right panel of Fig. \ref{fig:enty_prof} and they have been normalized by the predicted self-similar entropy. We note that we define entropy as
\begin{equation}
 K_{\Delta}\equiv\frac{k_{\rm{B}}T_{\Delta}}{n^{2/3}_{e,\Delta}}\,
\end{equation}
where $n_e$ is the electron number density and $\Delta$ is the chosen overdensity relative to the critical density of the Universe. At $z=0$ the the combined sample shows a higher normalization compared to the hot sample and its relaxed subset. This is due to it lower density profile and higher temperature profile. The gradients of the hot sample and the relaxed subset profiles steepen in the centre due to the accretion of low entropy gas. We compare with the observed median profiles of \citet{Pratt2010} and \citet{Giles2015}, and the baseline profile of \citet{VoitKayBryan2005} derived from non-radiative SPH simulations. The combined sample is in good agreement with the observations and tends to the non-radiative predictions at large radii.

At $z=1$ the three samples are in reasonable agreement with each other, all having a similar shape with the hot sample showing a marginally lower normalization. This change from $z=0$ is in agreement with the evolution in their density and temperature profiles. We compare the profiles to the observations of \citet{McDonald2014}. The combined and relaxed hot sample show good agreement with the observed profile for $r<0.5r_{500,\rm{spec}}$, but over predict the entropy at larger radii. In contrast the median profile of the hot sample is consistent with the observations at large radii, but under predicts the entropy in the centre of the cluster.

The departure from self-similarity for the three samples is due to a combination of the evolution in their temperature and density profiles. The relaxed hot sample shows a mild increase in entropy from $z=1$ to $z=0$ at large radii, due to change in its temperature profile, and a decrease in entropy in the core due to the increase in density at $z=0$. The increased normalization of the hot sample's temperature profile at $z=0$ compared to $z=1$ leads to an increased entropy profile with decreasing redshift, except in the core. The combined sample shows an increase in entropy at all radii at $z=0$ compared to $z=1$ and is produced by the decreased density and increased temperature with decreasing redshift.

\section{Summary \& Discussion}
\label{sec:sad}
In this work we have presented the MACSIS clusters, a sample of 390 zoomed simulations of the most massive and rarest clusters run with the state-of-the-art, calibrated baryonic physics model from the BAHAMAS project \citep{McCarthy2016} that yields realistic clusters. Such massive clusters are absent from the BAHAMAS simulation volumes of $596\,\mathrm{Mpc}$ as the simulated volume is too small. After introducing the selection of the sample from the parent $3.2\,\mathrm{Gpc}$ volume simulated with the \textit{Planck} 2013 cosmology, and demonstrating the agreement of the properties of our massive cluster sample with the properties of observed massive clusters, we examined the evolution of the cluster scaling relations and the evolution of the cluster gas profiles. 

By combining the MACSIS sample with the clusters in the BAHAMAS volume, we were able to examine the cluster scaling relations over the full observed mass range for the first time. Additionally, the MACSIS clusters enabled the study of the evolution of the cluster scaling relations to unprecedentedly high redshifts. Finally, the MACSIS sample enabled clusters to be selected in ways which mimic a cosmological study, such as selecting the hottest clusters, to examine if the scaling relations of such objects evolve differently from the underlying cluster population. Our main results are:
\begin{itemize}
    \item As shown in Fig. \ref{fig:observations}, the MACSIS simulations yield realistic massive clusters at low redshift and their progenitors are in good agreement with the limited observational data that is available at high redshift (i.e. $z=1$).
    \item Scaling relations for the combined sample that spans the full observed cluster mass range show significant deviations from the simple self-similar theory (see Figs. \ref{fig:MgMsr}-\ref{fig:LxTxsr}). Both the slope of the relations and the redshift evolution of the normalization are significantly affected by non-gravitational physics. The low redshift relations are in good agreement with observations and with most previous simulation work.
    \item The main drivers of non-self-similar evolution are AGN feedback, non-thermal pressure support and a mild mass dependence of the spectroscopic temperature bias. Shallower potentials of clusters that are less massive or form at lower redshifts allows feedback from AGN to eject more gas. Non-thermal pressure lowers a cluster's temperature for a given potential and is more important in more massive clusters that have had less time to thermalise. We found that the spectroscopic temperature bias increases for the most massive clusters.
    \item With the exception of the luminosity-temperature relation, we found the scatter about the best-fit scaling relations is insensitive to mass and redshift for all of the cluster samples.
    \item Selecting a hot cluster sample, i.e. core-excised spectroscopic temperatures $k_{\rm{B}}T^{\mathrm{X,ce}}_{500\mathrm{spec}}\geq5\,\rm{keV}$, significantly alters the scaling relations and their evolution. Excluding the spectroscopic temperature-total mass relation, we find that the scaling relations of the hot cluster sample evolve in a much more self-similar manner. After accounting for the expected self-similar evolution with redshift, we find that the normalizations are consistent with no evolution. The slopes of the best-fit relations at each redshift are also broadly consistent with the slopes predicted by self-similar theory. However, the spectroscopic temperature-total mass relation of the hot sample deviates further from self-similarity than the combined sample. Selecting hot clusters removes the less massive clusters from the sample, so the hot sample is dynamically younger than the combined sample as more massive clusters form later in the hierarchical merger scenario. This increases the average level of non-thermal support in the hot sample, leading to a flatter spectroscopic temperature-total mass relation. Additionally, the spectroscopic temperature bias flattens the relation for the most massive clusters and this has a larger impact in a sample of only hot clusters.
    \item Selecting a relaxed subset of hot clusters, where the most dynamically disturbed objects are removed, leads to a small reduction in the scatter for most scaling relations. Removing the most disturbed objects also leads to a reduction in the level of non-thermal support in the sample compared to the complete hot sample. This leads to steeper slope of the spectroscopic temperature-total mass relation compared to the hot sample and a value that is closer to the self-similar prediction.
    \item The median hot gas profiles of the combined sample in general shows good agreement with observed radial profiles. The low redshift data is in very good agreement, while the data at $z=1$ shows reasonable agreement with the relaxed hot sample.
    \item Comparison of the hot gas profiles at $z=0$ and $z=1$ show evolution different from self-similar prediction (see Figs. \ref{fig:gas_prof}-\ref{fig:enty_prof}). The combined sample shows a decreasing density profile with decreasing redshift, suggesting the impact of AGN feedback. Selecting a sample of hot clusters produces a median density profile that evolves in much more self-similar manner. The combined and hot samples have a median temperature profile that increases with decreasing redshift. This is likely driven by decreasing importance of non-thermal pressure support with decreasing redshift. Selecting relaxed hot cluster sample produces a median profile that evolves in better agreement with the self-similar prediction.
\end{itemize}
MACSIS enables the study of the observable properties of the most massive and rarest galaxy clusters. We have demonstrated that their progenitors provide a good match to the currently limited observational data at high redshift and that their observable properties evolve in a significantly more self-similar manner than for lower-mass and less-relaxed clusters. We have shown how the selection function can impact the derived scaling relations and radial profiles.

The size of the parent simulation enables the creation of synthetic lightcones with an area comparable to currently ongoing surveys. This will allow the impact of selection biases to be fully examined and the covariance of observable properties to be studied. Another route for future work is to improve our understanding of structure in the ICM, as the limited resolution and traditional SPH scheme used in this work limits our ability resolve structures and understand their impact on observable properties.

\section*{Acknowledgements}
This work used the DiRAC Data Centric system at Durham University, operated by the Institute for Computational Cosmology on behalf of the STFC DiRAC HPC Facility (www.dirac.ac.uk). This equipment was funded by BIS National E-infrastructure capital grant ST/K00042X/1, STFC capital grants ST/H008519/1 and ST/K00087X/1, STFC DiRAC Operations grant ST/K003267/1 and Durham University. DiRAC is part of the National E-Infrastructure. DJB and STK acknowledge support from STFC through grant ST/L000768/1. MAH is supported by an STFC quota studentship. IGM is supported by a STFC Advanced Fellowship. The research was supported in part by the European Research Council under the European Union's Seventh Framework Programme (FP7/2007-2013) / ERC Grant agreement 278594-GasAroundGalaxies. ARJ acknowledges support from STFC through grant ST/L00075X/1.

\bibliographystyle{mnras}
\bibliography{ms}

\appendix
\section{Selection Effects}
\label{app:seleff}
\begin{figure*}
 \includegraphics[width=\textwidth]{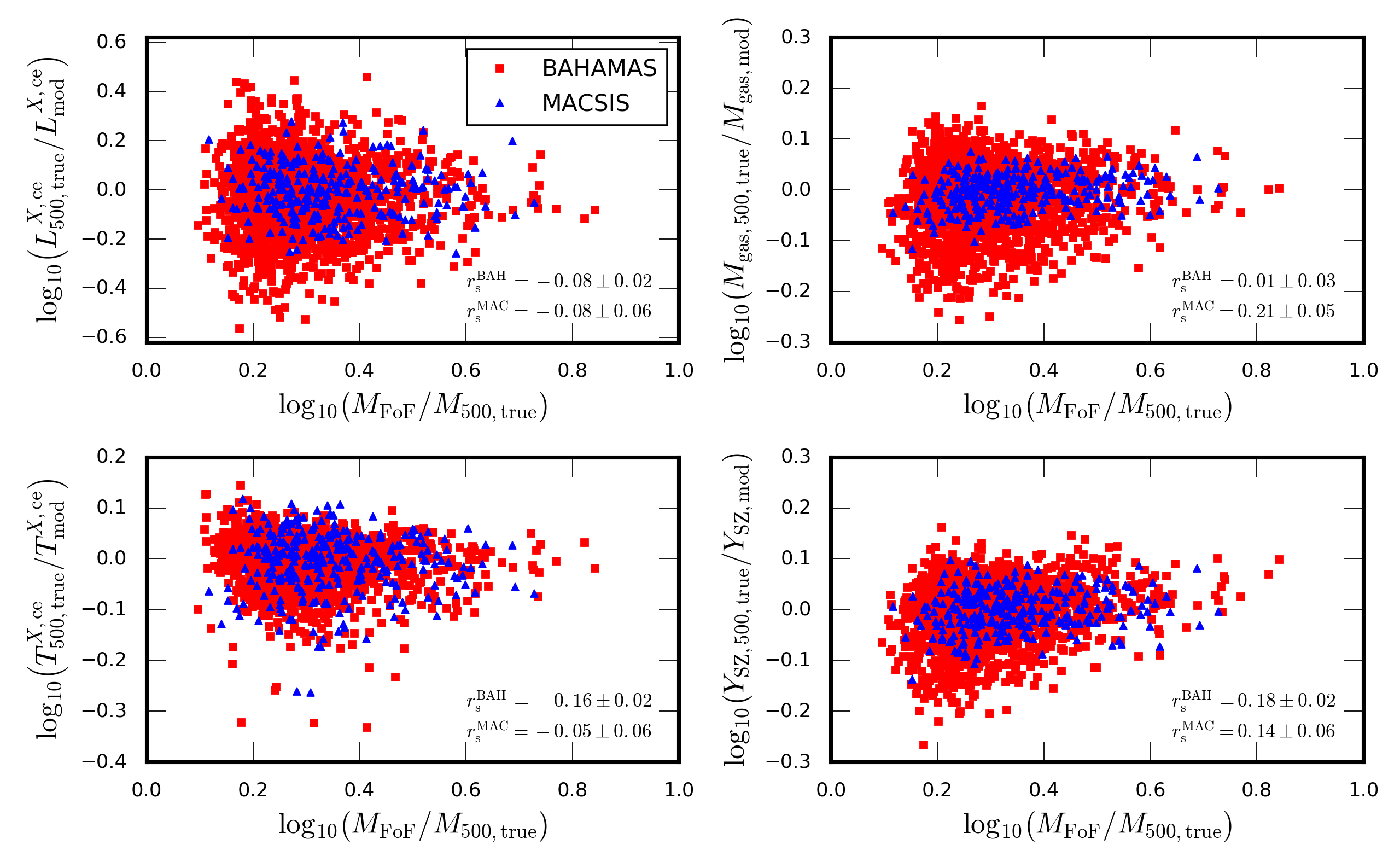}
 \caption{Plot showing the log of the ratio of the observable value and the value predicted by the best-fit power-law relation as a function of the log of the ratio of the cluster's $M_{\mathrm{FoF}}$ and $M_{500}$ for the bolometric X-ray luminosity (top left), gas mass (top right), spectroscopic temperature (bottom left) and integrated SZ effect (bottom right) for the BAHAMAS clusters (red squares) and the MACSIS clusters (blue triangles) that form the combined sample at $z=0$. The Spearman's rank correlation, $r_{\mathrm{s}}$, is shown for both samples with errors generated by bootstrapping the sample $10,000$ times. We do not find strong correlations for any of the ratios.}
 \label{fig:SelEff}
\end{figure*}

The selection of the MACSIS sample was done using the $M_{\mathrm{FoF}}$ mass of a halo in the parent simulation, but the scaling relations are presented using the $M_{500}$ cluster mass. This could potentially lead to a selection bias that impacts on the scaling relations presented in this work. Therefore, we make a mass cut to remove those clusters with large $M_{\rm{FoF}}/M_{500}$ ratio. To assess the impact of our sample selection we plot the log of the ratio of four cluster observables: the core-excised bolometric X-ray luminosity, the core-excised spectroscopic temperature, the gas mass and the integrated SZ signal, against their expected value from the best-fit relation as a function of the log of the ratio of the cluster's $M_{\rm{FoF}}$ and $M_{500}$ for the combined sample of clusters, with the cut included. Fig. \ref{fig:SelEff} shows the result of these plots. Any differences in the correlations of these ratios between MACSIS and BAHAMAS could indicate that selection effects were impacting the scaling relations. We have calculated the Spearman's rank correlation coefficients for both samples for all four observable ratios and find that only the gas mass shows a significant $(>2\sigma)$ difference between the two samples. However, all of the quantities show only weak correlations. We therefore conclude that the cut to remove clusters with extremely high $M_{\rm{FoF}}/M_{500}$ ratios has minimised any bias due to selection by $M_{\mathrm{FoF}}$.

\section{Self-similar relations}
\label{app:ssr}
If galaxy clusters were to form from a purely monolithic gravitational collapse, astrophysical processes were negligible and they were virialised, then we would expect them to be self-similar objects. This would mean that their properties would depend only on their mass \citep{WhiteRees1978,Kaiser1986}. The critical density of the Universe is defined as 
\begin{equation}
 \rho_{\rm{crit}}(z)\equiv E^2(z)\frac{3H^2_0}{8\pi G}\,,
\end{equation}
where $H_0$ is the Hubble constant, $G$ is the gravitational constant and
\begin{equation}
 E(z)\equiv\frac{H(z)}{H_0}=\sqrt{\Omega_{m}(1+z)^3+\Omega_{\Lambda}}\,.
\end{equation}
A cluster can then be defined as an overdensity with mass, $M$, inside a sphere of radius, $r$, with some average density, $\Delta$, relative to the critical density
\begin{equation}
 M_{\Delta}\propto\Delta\rho_{\rm{crit}}(z)r^3_{\Delta}\propto E^{2}(z)\,r^3_{\Delta}.
\end{equation}
As gas collapses into the potential, $\Phi$, of the cluster it is heated and, assuming that it is a collapsed iosthermal sphere, it will reach a temperature, $T$, of
\begin{equation}
 k_{B}T_{\Delta}\equiv\frac{1}{2}\Phi=\frac{GM_{\Delta}\mu m_{p}}{2r_{\Delta}}\,,
\end{equation}
where $k_{B}$ is the Boltzmann constant, $m_{p}$ is the mass of the proton and $\mu$ is the mean molecular weight. Therefore, the self-similar temperature of the cluster is proportional to its mass via
\begin{equation} \label{eq:Tx-M}
 T_{\Delta}\propto M^{2/3}_{\Delta}E^{2/3}(z)\,.
\end{equation}
Under the assumption that main cooling mechanism of the cluster is thermal bremsstrahlung, the cluster gas will emit X-rays and its bolometric emission is is proportional to
\begin{equation}
 L_{\rm{X},\Delta}^{\rm{bol}}\propto\rho^2\Lambda(T)r_{\Delta}^3\propto\rho^2T^{1/2}r_{\Delta}^3\propto M^{4/3}_{\Delta}E^{7/3}(z)\,,
\end{equation}
where the cooling function $\Lambda(T)\propto T^{1/2}$ for the bolometric case \citep[e.g.][]{Sarazin1986}. Using equation (\ref{eq:Tx-M}), we can derive the self-similar prediction for the X-ray luminosity-temperature relation
\begin{equation}
 L_{\rm{X},\Delta}^{\rm{bol}}\propto T^{2}E(z)\,.
\end{equation}
Assuming a constant gas fraction, the integrated Sunyaev-Zel'dovich signal, $Y_{\rm{SZ}}$, and its X-ray analogue, $Y_{\rm{X}}$, of the cluster can be predicted by
\begin{equation}
 Y_{\rm{SZ},\Delta}\propto Y_{\rm{X},\Delta}\equiv M_{\Delta}T_{\Delta}\,,
\end{equation}
and the self-similar relations are
\begin{equation}
 Y_{\rm{SZ},\Delta}\propto M_{\Delta}^{5/3}E^{2/3}(z)\,,
\end{equation}
\begin{equation}
 Y_{\rm{X},\Delta}\propto M_{\Delta}^{5/3}E^{2/3}(z)\,.
\end{equation}

\section{Fit parameters}
\label{app:fitpar}
The tables \ref{tab:Lx-Mtab}-\ref{tab:Lx-Txtab} below list the parameter values for the best-fit relations of the scaling relations presented in this paper. For $z>1$ there are too few clusters in too many bins to reliably measure a bit-fit relation for the hot cluster sample and the relaxed subset and these values are not presented. 

\renewcommand\arraystretch{1.5}
\begin{table*}
 \caption{Normalization, slope and scatter about the best-fit bolometric luminosity-total mass relations for the three samples (see eq. \ref{eq:plfit}). All quantities presented in this table are `$\rm{spec}$' values calculated via the synthetic X-ray analysis.}
 \centering
 \begin{tabularx}{\textwidth}{l C C C C C C C C C}
  \hline
  Redshift & \multicolumn{3}{c}{Combined sample} & \multicolumn{3}{c}{Hot Clusters} & \multicolumn{3}{c}{Relaxed, Hot Clusters} \\
   & $A$ & $\alpha$ & $\langle\sigma_{\log_{10}Y}\rangle$ & $A$ & $\alpha$ & $\langle\sigma_{\log_{10}Y}\rangle$ & $A$ & $\alpha$ & $\langle\sigma_{\log_{10}Y}\rangle$ \\
  \hline
  $0.00$ & $44.50^{+0.01}_{-0.01}$ & $1.88^{+0.03}_{-0.05}$ & $0.15^{+0.01}_{-0.02}$ & $44.71^{+0.02}_{-0.02}$ & $1.36^{+0.08}_{-0.07}$ & $0.12^{+0.01}_{-0.02}$ & $44.69^{+0.03}_{-0.03}$ & $1.43^{+0.13}_{-0.09}$ & $0.11^{+0.01}_{-0.01}$ \\
  $0.25$ & $44.60^{+0.01}_{-0.02}$ & $1.98^{+0.03}_{-0.05}$ & $0.12^{+0.01}_{-0.02}$ & $44.74^{+0.03}_{-0.03}$ & $1.42^{+0.14}_{-0.13}$ & $0.12^{+0.02}_{-0.01}$ & $44.69^{+0.03}_{-0.03}$ & $1.58^{+0.10}_{-0.13}$ & $0.09^{+0.01}_{-0.01}$ \\
  $0.50$ & $44.63^{+0.01}_{-0.01}$ & $1.91^{+0.03}_{-0.04}$ & $0.11^{+0.01}_{-0.01}$ & $44.74^{+0.02}_{-0.02}$ & $1.32^{+0.10}_{-0.11}$ & $0.12^{+0.01}_{-0.02}$ & $44.73^{+0.01}_{-0.01}$ & $1.44^{+0.13}_{-0.09}$ & $0.10^{+0.01}_{-0.03}$ \\
  $1.00$ & $44.75^{+0.08}_{-0.06}$ & $2.02^{+0.19}_{-0.14}$ & $0.12^{+0.01}_{-0.02}$ & $-$ & $-$ & $-$ & $-$ & $-$ & $-$ \\
  $1.50$ & $44.98^{+0.19}_{-0.12}$ & $2.13^{+0.32}_{-0.21}$ & $0.13^{+0.01}_{-0.01}$ & $-$ & $-$ & $-$ & $-$ & $-$ & $-$ \\
  \hline
 \end{tabularx}
 \label{tab:Lx-Mtab}
\end{table*}

\begin{table*}
 \caption{Normalization, slope and scatter about the best-fit spectroscopic temperature-total mass relations for the three samples (see eq. \ref{eq:plfit}). All quantities presented in this table are `$\rm{spec}$' values calculated via the synthetic X-ray analysis.}
 \centering
 \begin{tabularx}{\textwidth}{l C C C C C C C C C}
  \hline
  Redshift & \multicolumn{3}{c}{Combined sample} & \multicolumn{3}{c}{Hot Clusters} & \multicolumn{3}{c}{Relaxed, Hot Clusters} \\
   & $A$ & $\alpha$ & $\langle\sigma_{\log_{10}Y}\rangle$ & $A$ & $\alpha$ & $\langle\sigma_{\log_{10}Y}\rangle$ & $A$ & $\alpha$ & $\langle\sigma_{\log_{10}Y}\rangle$ \\
  \hline
  $0.00$ & $0.68^{+0.00}_{-0.00}$ & $0.58^{+0.01}_{-0.01}$ & $0.05^{+0.003}_{-0.003}$ & $0.71^{+0.01}_{-0.01}$ & $0.51^{+0.04}_{-0.04}$ & $0.05^{+0.01}_{-0.002}$ & $0.70^{+0.01}_{-0.01}$ & $0.55^{+0.06}_{-0.03}$ & $0.04^{+0.003}_{-0.010}$ \\
  $0.25$ & $0.67^{+0.00}_{-0.01}$ & $0.60^{+0.01}_{-0.01}$ & $0.04^{+0.004}_{-0.001}$ & $0.69^{+0.01}_{-0.01}$ & $0.50^{+0.07}_{-0.05}$ & $0.04^{+0.005}_{-0.002}$ & $0.68^{+0.01}_{-0.01}$ & $0.58^{+0.05}_{-0.06}$ & $0.04^{+0.005}_{-0.005}$ \\
  $0.50$ & $0.64^{+0.01}_{-0.01}$ & $0.57^{+0.02}_{-0.01}$ & $0.04^{+0.002}_{-0.001}$ & $0.66^{+0.01}_{-0.01}$ & $0.46^{+0.07}_{-0.05}$ & $0.05^{+0.008}_{-0.012}$ & $0.66^{+0.01}_{-0.01}$ & $0.51^{+0.05}_{-0.07}$ & $0.03^{+0.009}_{-0.004}$ \\
  $1.00$ & $0.61^{+0.02}_{-0.02}$ & $0.58^{+0.04}_{-0.05}$ & $0.05^{+0.001}_{-0.003}$ & $-$ & $-$ & $-$ & $-$ & $-$ & $-$ \\
  $1.50$ & $0.60^{+0.03}_{-0.03}$ & $0.61^{+0.06}_{-0.06}$ & $0.04^{+0.001}_{-0.002}$ & $-$ & $-$ & $-$ & $-$ & $-$ & $-$ \\
  \hline
 \end{tabularx}
 \label{tab:Tx-Mtab}
\end{table*}

\begin{table*}
 \caption{Normalization, slope and scatter about the best-fit gas mass-total mass relations for the three samples (see eq. \ref{eq:plfit}). All quantities presented in this table are `$\rm{spec}$' values calculated via the synthetic X-ray analysis.}
 \centering
 \begin{tabularx}{\textwidth}{l C C C C C C C C C}
  \hline
  Redshift & \multicolumn{3}{c}{Combined sample} & \multicolumn{3}{c}{Hot Clusters} & \multicolumn{3}{c}{Relaxed, Hot Clusters} \\
   & $A$ & $\alpha$ & $\langle\sigma_{\log_{10}Y}\rangle$ & $A$ & $\alpha$ & $\langle\sigma_{\log_{10}Y}\rangle$ & $A$ & $\alpha$ & $\langle\sigma_{\log_{10}Y}\rangle$ \\
  \hline
  $0.00$ & $13.67^{+0.01}_{-0.01}$ & $1.25^{+0.01}_{-0.03}$ & $0.07^{+0.01}_{-0.01}$ & $13.77^{+0.01}_{-0.01}$ & $1.02^{+0.03}_{-0.03}$ & $0.06^{+0.01}_{-0.01}$ & $13.75^{+0.01}_{-0.01}$ & $1.05^{+0.04}_{-0.04}$ & $0.05^{+0.01}_{-0.01}$ \\
  $0.25$ & $13.72^{+0.00}_{-0.01}$ & $1.29^{+0.01}_{-0.02}$ & $0.06^{+0.01}_{-0.01}$ & $13.79^{+0.01}_{-0.01}$ & $1.04^{+0.04}_{-0.06}$ & $0.06^{+0.01}_{-0.01}$ & $13.77^{+0.01}_{-0.01}$ & $1.09^{+0.04}_{-0.04}$ & $0.04^{+0.01}_{-0.01}$ \\
  $0.50$ & $13.73^{+0.01}_{-0.01}$ & $1.25^{+0.03}_{-0.02}$ & $0.07^{+0.01}_{-0.01}$ & $13.80^{+0.01}_{-0.01}$ & $0.92^{+0.06}_{-0.05}$ & $0.05^{+0.01}_{-0.01}$ & $13.79^{+0.01}_{-0.01}$ & $0.97^{+0.08}_{-0.06}$ & $0.04^{+0.01}_{-0.01}$ \\
  $1.00$ & $13.77^{+0.04}_{-0.03}$ & $1.29^{+0.09}_{-0.07}$ & $0.07^{+0.01}_{-0.01}$ & $-$ & $-$ & $-$ & $-$ & $-$ & $-$ \\
  $1.50$ & $13.85^{+0.05}_{-0.08}$ & $1.31^{+0.11}_{-0.14}$ & $0.06^{+0.01}_{-0.01}$ & $-$ & $-$ & $-$ & $-$ & $-$ & $-$ \\
  \hline
 \end{tabularx}
 \label{tab:Mg-Mtab}
\end{table*}

\begin{table*}
 \caption{Normalization, slope and scatter about the best-fit X-ray analogue $Y$-total mass relations for the three samples (see eq. \ref{eq:plfit}). All quantities presented in this table are `$\rm{spec}$' values calculated via the synthetic X-ray analysis.}
 \centering
 \begin{tabularx}{\textwidth}{l C C C C C C C C C}
  \hline
  Redshift & \multicolumn{3}{c}{Combined sample} & \multicolumn{3}{c}{Hot Clusters} & \multicolumn{3}{c}{Relaxed, Hot Clusters} \\
   & $A$ & $\alpha$ & $\langle\sigma_{\log_{10}Y}\rangle$ & $A$ & $\alpha$ & $\langle\sigma_{\log_{10}Y}\rangle$ & $A$ & $\alpha$ & $\langle\sigma_{\log_{10}Y}\rangle$ \\
  \hline
  $0.00$ & $14.33^{+0.01}_{-0.01}$ & $1.84^{+0.02}_{-0.05}$ & $0.12^{+0.01}_{-0.01}$ & $14.47^{+0.02}_{-0.02}$ & $1.51^{+0.07}_{-0.08}$ & $0.11^{+0.01}_{-0.01}$ & $14.45^{+0.02}_{-0.02}$ & $1.59^{+0.12}_{-0.06}$ & $0.08^{+0.01}_{-0.01}$ \\
  $0.25$ & $14.38^{+0.01}_{-0.01}$ & $1.91^{+0.02}_{-0.04}$ & $0.11^{+0.01}_{-0.01}$ & $14.47^{+0.02}_{-0.02}$ & $1.57^{+0.09}_{-0.12}$ & $0.10^{+0.01}_{-0.01}$ & $14.45^{+0.02}_{-0.02}$ & $1.67^{+0.09}_{-0.08}$ & $0.07^{+0.01}_{-0.01}$ \\
  $0.50$ & $14.37^{+0.01}_{-0.01}$ & $1.85^{+0.04}_{-0.04}$ & $0.11^{+0.01}_{-0.01}$ & $14.47^{+0.02}_{-0.01}$ & $1.35^{+0.10}_{-0.09}$ & $0.10^{+0.01}_{-0.01}$ & $14.45^{+0.02}_{-0.01}$ & $1.45^{+0.15}_{-0.11}$ & $0.07^{+0.01}_{-0.01}$ \\
  $1.00$ & $14.39^{+0.07}_{-0.05}$ & $1.89^{+0.15}_{-0.12}$ & $0.12^{+0.01}_{-0.01}$ & $-$ & $-$ & $-$ & $-$ & $-$ & $-$ \\
  $1.50$ & $14.48^{+0.08}_{-0.06}$ & $1.98^{+0.18}_{-0.13}$ & $0.10^{+0.01}_{-0.01}$ & $-$ & $-$ & $-$ & $-$ & $-$ & $-$ \\
  \hline
 \end{tabularx}
 \label{tab:Yx-Mtab}
\end{table*}

\begin{table*}
 \caption{Normalization, slope and scatter about the best-fit integrated SZ signal-total mass relations for the three samples (see eq. \ref{eq:plfit}). All quantities presented in this table are `$\rm{spec}$' values calculated via the synthetic X-ray analysis.}
 \centering
 \begin{tabularx}{\textwidth}{l C C C C C C C C C}
  \hline
  Redshift & \multicolumn{3}{c}{Combined sample} & \multicolumn{3}{c}{Hot Clusters} & \multicolumn{3}{c}{Relaxed, Hot Clusters} \\
   & $A$ & $\alpha$ & $\langle\sigma_{\log_{10}Y}\rangle$ & $A$ & $\alpha$ & $\langle\sigma_{\log_{10}Y}\rangle$ & $A$ & $\alpha$ & $\langle\sigma_{\log_{10}Y}\rangle$ \\
  \hline
  $0.00$ & $-4.51^{+0.01}_{-0.01}$ & $1.88^{+0.02}_{-0.03}$ & $0.10^{+0.01}_{-0.01}$ & $-4.39^{+0.02}_{-0.02}$ & $1.60^{+0.07}_{-0.05}$ & $0.10^{+0.01}_{-0.02}$ & $-4.42^{+0.02}_{-0.02}$ & $1.69^{+0.07}_{-0.07}$ & $0.09^{+0.01}_{-0.01}$ \\
  $0.25$ & $-4.46^{+0.01}_{-0.01}$ & $1.94^{+0.02}_{-0.03}$ & $0.10^{+0.01}_{-0.01}$ & $-4.36^{+0.02}_{-0.02}$ & $1.62^{+0.10}_{-0.11}$ & $0.10^{+0.01}_{-0.01}$ & $-4.40^{+0.02}_{-0.03}$ & $1.74^{+0.09}_{-0.09}$ & $0.08^{+0.01}_{-0.01}$ \\
  $0.50$ & $-4.45^{+0.01}_{-0.01}$ & $1.88^{+0.03}_{-0.03}$ & $0.10^{+0.01}_{-0.01}$ & $-4.37^{+0.02}_{-0.02}$ & $1.48^{+0.10}_{-0.10}$ & $0.10^{+0.01}_{-0.01}$ & $-4.38^{+0.02}_{-0.01}$ & $1.59^{+0.17}_{-0.14}$ & $0.08^{+0.01}_{-0.01}$ \\
  $1.00$ & $-4.41^{+0.07}_{-0.05}$ & $1.91^{+0.15}_{-0.11}$ & $0.11^{+0.01}_{-0.01}$ & $-$ & $-$ & $-$ & $-$ & $-$ & $-$ \\
  $1.50$ & $-4.29^{+0.05}_{-0.06}$ & $2.04^{+0.09}_{-0.12}$ & $0.10^{+0.01}_{-0.01}$ & $-$ & $-$ & $-$ & $-$ & $-$ & $-$ \\
  \hline
 \end{tabularx}
 \label{tab:Ysz-Mtab}
\end{table*}

\begin{table*}
 \caption{Normalization, slope and scatter about the best-fit bolometric luminosity-spectroscopic temperature relations for the three samples (see eq. \ref{eq:plfit}). All quantities presented in this table are `$\rm{spec}$' values calculated via the synthetic X-ray analysis.}
 \centering
 \begin{tabularx}{\textwidth}{l C C C C C C C C C}
  \hline
  Redshift & \multicolumn{3}{c}{Combined sample} & \multicolumn{3}{c}{Hot Clusters} & \multicolumn{3}{c}{Relaxed, Hot Clusters} \\
   & $A$ & $\alpha$ & $\langle\sigma_{\log_{10}Y}\rangle$ & $A$ & $\alpha$ & $\langle\sigma_{\log_{10}Y}\rangle$ & $A$ & $\alpha$ & $\langle\sigma_{\log_{10}Y}\rangle$ \\
  \hline
  $0.00$ & $44.80^{+0.02}_{-0.01}$ & $3.01^{+0.04}_{-0.04}$ & $0.14^{+0.01}_{-0.01}$ & $44.93^{+0.01}_{-0.01}$ & $2.41^{+0.12}_{-0.12}$ & $0.11^{+0.01}_{-0.01}$ & $44.89^{+0.02}_{-0.02}$ & $2.53^{+0.12}_{-0.13}$ & $0.10^{+0.01}_{-0.01}$ \\
  $0.25$ & $44.89^{+0.01}_{-0.01}$ & $3.15^{+0.03}_{-0.04}$ & $0.12^{+0.01}_{-0.01}$ & $44.95^{+0.02}_{-0.01}$ & $2.82^{+0.16}_{-0.21}$ & $0.11^{+0.01}_{-0.01}$ & $44.94^{+0.02}_{-0.02}$ & $2.67^{+0.16}_{-0.17}$ & $0.09^{+0.01}_{-0.01}$ \\
  $0.50$ & $44.94^{+0.01}_{-0.01}$ & $3.19^{+0.03}_{-0.03}$ & $0.11^{+0.01}_{-0.01}$ & $44.99^{+0.01}_{-0.01}$ & $2.67^{+0.12}_{-0.19}$ & $0.10^{+0.01}_{-0.01}$ & $44.97^{+0.01}_{-0.02}$ & $2.62^{+0.27}_{-0.17}$ & $0.08^{+0.01}_{-0.01}$ \\
  $1.00$ & $45.08^{+0.02}_{-0.02}$ & $3.36^{+0.05}_{-0.08}$ & $0.13^{+0.01}_{-0.01}$ & $-$ & $-$ & $-$ & $-$ & $-$ & $-$ \\
  $1.50$ & $45.19^{+0.13}_{-0.11}$ & $3.45^{+0.37}_{-0.31}$ & $0.12^{+0.01}_{-0.01}$ & $-$ & $-$ & $-$ & $-$ & $-$ & $-$ \\
  \hline
 \end{tabularx}
 \label{tab:Lx-Txtab}
\end{table*}
\renewcommand\arraystretch{1.0}

\bsp
\label{lastpage}
\end{document}